\documentclass[
 reprint,
 superscriptaddress,
 nofootinbib,
 amsmath,amssymb,
 aps,
]{revtex4-1}

\usepackage{graphicx}
\usepackage{dcolumn}
\usepackage{bm}
\usepackage{color}
\usepackage[usenames,dvipsnames]{xcolor}
\usepackage[breaklinks, pdftex, hyperfootnotes=true, pdfpagelabels, bookmarks, pageanchor]{hyperref}
\usepackage{ulem}
\hypersetup{
	colorlinks=true, linktocpage=true, pdfstartpage=1, pdfstartview=FitH, pdfborder={0 0 0},
	breaklinks=true, pdfpagemode=UseNone, pageanchor=true, pdfpagemode=UseOutlines,
	plainpages=false, bookmarksnumbered, bookmarksopen=true, bookmarksopenlevel=1,
	hypertexnames=true, pdfhighlight=/O,
	urlcolor=blue, linkcolor=RubineRed, citecolor=blue,
}

\newcommand{\bes} {\begin{subequations}}
\newcommand{\ees} {\end{subequations}}

\begin{document}

\title{Probing the Universality of Topological Defect Formation in a Quantum Annealer: Kibble-Zurek Mechanism and Beyond}

\author{Yuki Bando}
\email{y-bando@qa.iir.titech.ac.jp}
\affiliation{Institute of Innovative Research, Tokyo Institute of Technology, Nagatsuta-cho, Midori-ku, Yokohama 226-8503, Japan}

\author{Yuki Susa}
\thanks{Present address: System Platform Research Laboratories, NEC Corporation, Kawasaki 211-8666, Japan}
\affiliation{Institute of Innovative Research, Tokyo Institute of Technology, Oh-okayama, Meguro-ku, Tokyo 152-8550, Japan}

\author{Hiroki Oshiyama}
\affiliation{Department of Physics, Tohoku University, Sendai 980-8578, Japan}

\author{Naokazu Shibata}
\affiliation{Department of Physics, Tohoku University, Sendai 980-8578, Japan}
\email{}
\author{Masayuki Ohzeki}
\affiliation{Graduate School of Information Sciences, Tohoku University, Sendai 980-8579, Japan}
\affiliation{Institute of Innovative Research, Tokyo Institute of Technology, Nagatsuta-cho, Midori-ku, Yokohama 226-8503, Japan}
\affiliation{Sigma-i Co. Ltd., Konan, Minato-ku, Tokyo 108-0075, Japan}

\author{Fernando Javier G\'omez-Ruiz}
\affiliation{Donostia International Physics Center, E-20018 San Sebasti\'an, Spain}

\author{Daniel A. Lidar}
\affiliation{Departments of Electrical and Computer Engineering, Chemistry, and Physics, University of Southern California, Los Angeles, CA 90089, USA}
\affiliation{Center for Quantum Information Science \& Technology, University of Southern California, Los Angeles, CA 90089, USA}

\author{Adolfo del Campo}
\affiliation{Donostia International Physics Center, E-20018 San Sebasti\'an, Spain}
\affiliation{IKERBASQUE, Basque Foundation for Science, E-48013 Bilbao, Spain}
\affiliation{Department of Physics, University of Massachusetts Boston,100 Morrissey Boulevard, Boston, MA 02125, USA}

\author{Sei Suzuki}
\affiliation{Department of Liberal Arts, Saitama Medical University, Moroyama, Saitama 350-0495, Japan}

\author{Hidetoshi Nishimori}
\affiliation{Institute of Innovative Research, Tokyo Institute of Technology, Nagatsuta-cho, Midori-ku, Yokohama 226-8503, Japan}
\affiliation{Graduate School of Information Sciences, Tohoku University, Sendai 980-8579, Japan}
\affiliation{RIKEN, Interdisciplinary Theoretical and Mathematical Sciences (iTHEMS),
Wako, Saitama 351-0198, Japan}

\date{\today}

\begin{abstract}
The number of topological defects created in a system driven through a quantum phase transition exhibits a power-law scaling with the driving time. This universal scaling law is the key prediction of the Kibble-Zurek mechanism (KZM), and testing it using a hardware-based quantum simulator is a coveted goal of quantum information science. Here we provide such a test using quantum annealing. Specifically, we report on extensive experimental tests of topological defect formation via the one-dimensional transverse-field Ising model on two different D-Wave quantum annealing devices.
We find that the quantum simulator results can indeed be explained by the KZM for open-system quantum dynamics with phase-flip errors, with certain quantitative deviations from the theory likely caused by factors such as random control errors and transient effects.
In addition, we probe physics beyond the KZM by identifying signatures of universality in the distribution and cumulants of the number of kinks and their decay, and again find agreement with the quantum simulator results. 
This implies that the theoretical predictions of the generalized KZM theory, which assumes isolation from the environment, applies beyond its original scope to an open system. We support this result by extensive numerical computations.
To check whether an alternative, classical interpretation of these results is possible, we used the spin-vector Monte Carlo model, a candidate classical description of the D-Wave device.
We find that the degree of agreement with the experimental data from the D-Wave annealing devices is better for the KZM, a quantum theory, than for the classical spin-vector Monte Carlo model, thus favoring a quantum description of the device.
Our work provides an experimental test of quantum critical dynamics in an open quantum system, and paves the way to new directions in quantum simulation experiments.
\end{abstract}

\keywords{Quantum annealing, Kibble-Zurek mechanism, Transverse-field Ising model}

\maketitle

\section{\label{sec:level1}Introduction}

Quantum simulations are emerging to be one of the important applications of quantum annealing~\cite{kadowaki_quantum_1998,RevModPhys.80.1061,Albash-Lidar:RMP,Hauke:2019aa}, quite different, and arguably more natural, than the original intent of using such devices for optimization, the subject of many recent studies~\cite{q108,speedup,Hen:2015rt,DW2000Q,King:2015cs,PhysRevX.6.031015,2016arXiv160401746M,Vinci:2016tg,Albash:2017aa,Mandra:2017ab,Pearson:2019aa}. Prominent examples include the simulation of the Kosterlitz-Thouless topological phase transition~\cite{King2018,King2019} and three-dimensional spin glasses~\cite{Harris2018} using the D-Wave quantum annealing devices, that have successfully reproduced the behavior of various physical quantities and the structure of the phase diagram, as predicted by classical simulations. Quantum simulation has also been pursued using other systems such as trapped ions~\cite{Islam:2013mi,Smith:2016aa,Zhang:2017aa}.

Here we use D-Wave quantum annealers to perform quantum simulations of the Kibble-Zurek mechanism (KZM)~\cite{Kibble1976,Zurek1985}, which predicts the kink (or defect~\cite{kink2defect}) formation when a system crosses a phase transition point at a finite rate.  While the theory of the KZM was originally formulated for classical phase transitions, it has been extended to describe quantum critical dynamics~\cite{Polkovnikov05,Damski05,Dziarmaga05,Zurek2005}. As the dominant paradigm to describe the universal dynamics of a quantum phase transition, it has motivated a wide variety of experimental and theoretical studies~\cite{Polkovnikov11,DZ14,Uwe07,Uwe10}.
Laboratory tests of the KZM in quantum platforms have been carried out pursuing different quantum simulation approaches, e.g., using qubits to emulate free fermion models~\cite{Xu2014,Cui16,Wu16,Cui19} and via a fully digital approach using Rydberg atoms~\cite{Lukin17}.

Tests of the KZM can also be used to quantitatively assess the performance of a quantum device. Indeed, the KZM scaling is sensitive to, e.g., nonlinear driving schemes~\cite{Diptiman08,Barankov08}, disorder~\cite{Dziarmaga2006,Caneva07}, inhomogeneities in the system~\cite{ZD08,DM10,DKZ13,Fernando19} and decoherence~\cite{Patane08,Nalbach15,Dutta16,puebla2019akzm}.
The use of the KZM to assess the performance of a quantum annealer was studied by Gardas \textit{et al}.~\cite{Gardas2018}, focusing on the one-dimensional case on two previous generation D-Wave 2X devices, and by Weinberg \textit{et al}.~\cite{Weinberg2019} on a current generation D-Wave 2000Q (DW2KQ) device, focusing on the Ising Hamiltonian on the two-dimensional square lattice.

Here we report on extensive DW2KQ experiments for the one-dimensional transverse-field Ising model, using two separate realizations of the device to perform quantum simulations of the predictions of the KZM for the kink density.  We also test a recent theory for the kink density distribution developed by one of us~\cite{DelCampo2018}, thus probing physics beyond the original KZM prediction of the average number of kinks~\cite{Polkovnikov05,Damski05,Dziarmaga05,Zurek2005}. Unlike Gardas \textit{et al}.~\cite{Gardas2018}, our work finds a universal power-law scaling behavior of the average number of kinks. We choose the one-dimensional problem because departures from the ideal theoretical setting due to noise and other reasons would easily destroy ordering in one dimension, and therefore it is easy to detect the effects of imperfections in one dimension, implying  that the data would clearly reveal open system effects.  It is also an advantage of the one-dimensional problem that we can avoid the problem of embedding of the system on the Chimera graph of the D-Wave device~\cite{Choi2008}. Moreover, previous studies of both antiferromagnetic~\cite{PAL:13,Mishra:2015} and ferromagnetic chains~\cite{Mishra2018} using previous generations of D-Wave devices obtained good agreement with open quantum systems theory~\cite{Breuer:book}.

Our work establishes the power-law scaling behavior of the average number of kinks, variance and third-order cumulant with the timescale in which the transition is crossed. In doing so, we provide a strategy to assess the behavior of quantum annealers, and find it to be well described within the framework of open system quantum dynamics. Our work thus provides an experimental test of quantum critical dynamics in an open quantum system. The universal power law scaling found in the cumulants of the kink-number distribution  shows that signatures of universality beyond the KZM recently predicted in isolated quantum critical systems continue to hold in the presence of coupling to an environment, to which we provide support by numerical computations.

This paper is organized as follows. Background on the KZM and its generalization, the problem we study, and the experimental methods are described in Sec.~\ref{sec:Problem}.  The empirical results on the kink density are presented and compared with the generalized KZM theory in Sec.~\ref{sec:pl}, and Sec.~\ref{sec:dist} similarly presents the kink distribution results.
In Sec.~\ref{section:classical} we address the question of whether classical models suffice to explain our empirical results. We do this by modeling the kink distribution using the classical Boltzmann distribution of the Ising spin chain, and by comparing the empirical results to the predictions of the classical spin-vector Monte Carlo model. We close the paper with a discussion in Sec.~\ref{sec:discussion}, including a comparison with Refs.~\cite{Gardas2018,Weinberg2019}, and conclude in Sec.~\ref{sec:summary}. Additional materials are presented in the Appendixes.

\section{Theoretical and experimental background}
\label{sec:Problem}

We first describe the problem to be studied, and then explain the predictions of the KZM, followed by our experimental methods for testing the theoretical predictions.

\subsection{The problem studied}

\begin{figure}[b]
\includegraphics[width=\columnwidth]{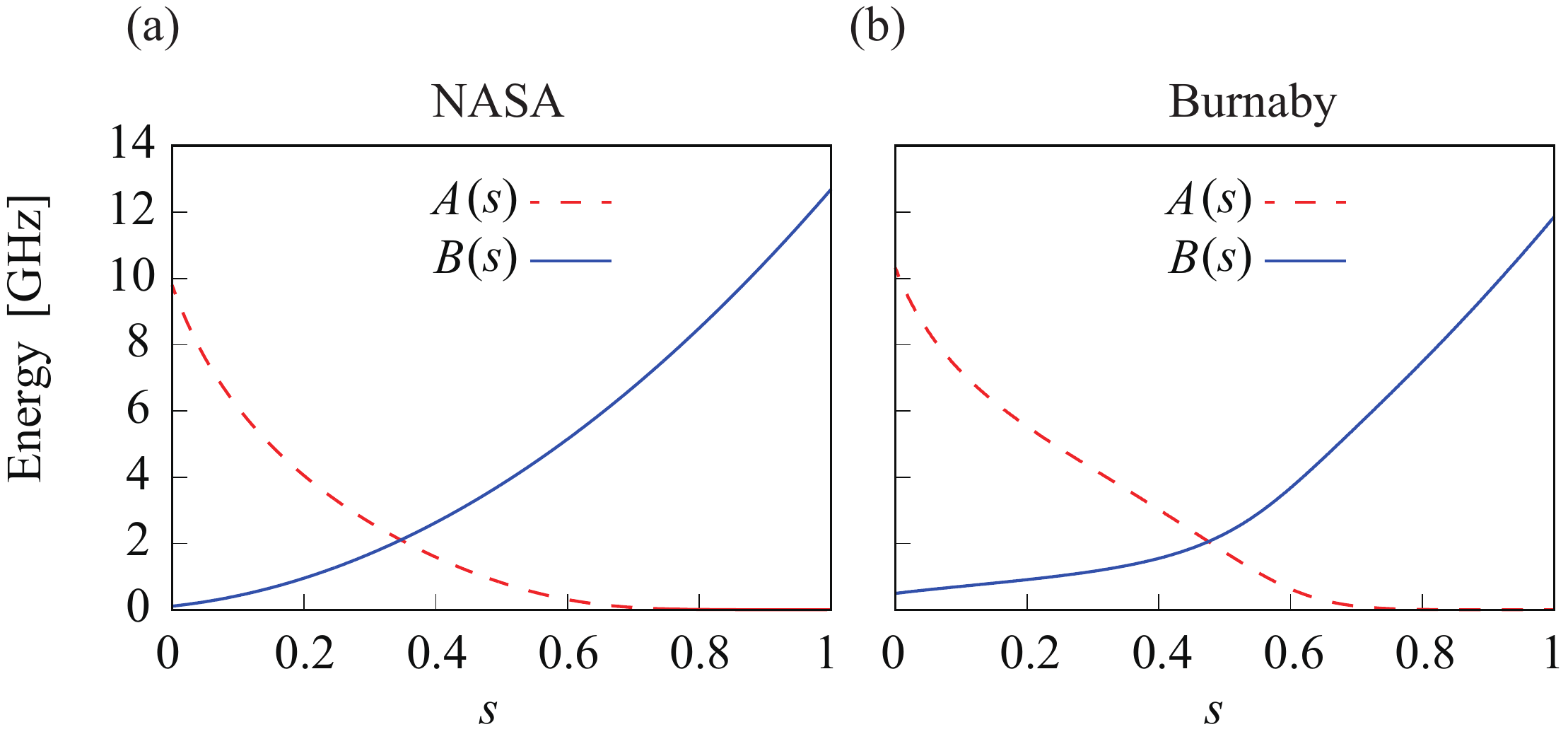}
\caption{
The annealing schedules on the DW2KQ quantum annealers at (a) NASA Ames Research Center and (b) Burnaby. For the actual quantum annealing processes, $A(s)/2$ and $B(s)/2$ are used as in Eq.~\eqref{eq:original_Hamiltonian}. The energy scale is converted into frequency, i.e., the vertical axis is $E/h$, where $h$ is the Planck constant.}
\label{fig:schedule}
\end{figure}

The target system of our investigation is the one-dimensional transverse-field Ising model defined and parameterized by the Hamiltonian,
\begin{equation}
    H(s)=\frac{A(s)}{2}\sum_{i=1}^L \sigma_i^{x}+\frac{B(s)}{2}\sum_{i=1}^{L-1} J\sigma_i^z \sigma_{i+1}^z,
    \label{eq:original_Hamiltonian}
\end{equation}
where $L$ is the chain length (system size), $s=t/t_a$, with the time $t\in[0,t_a]$ and the final time $t=t_a$ being the annealing time.
The absolute value of the interaction strength is chosen to be $|J|=1$, and as explained below we consider both the ferromagnetic ($J<0$) and antiferromagnetic ($J>0$) cases. We adopt a free boundary condition as indicated by the upper bound $L-1$ in Eq.~\eqref{eq:original_Hamiltonian}. 
Effects of the choice of a specific boundary condition are of the order of $1/L$, which is much smaller than the statistical fluctuations in the data shown below.
The functional forms of the annealing schedules $A(s)$ and $B(s)$ are shown in Fig.~\ref{fig:schedule}. 

This one-dimensional model has a second-order quantum phase transition at $A(s)=B(s)$ for a time-independent system, i.e., when $s$ is regarded as a fixed parameter~\cite{Nishimoribook2011}. The system is in the ferromagnetic phase when $A(s)<B(s)$ and is paramagnetic for $A(s)>B(s)$. The system is initially in the paramagnetic phase since $A(0)>0$ and $B(0)\approx 0$ as seen in Fig.~\ref{fig:schedule}. The rate of change of the annealing schedules $A(s)$ and $B(s)$ is finite, and the system does not necessarily follow the instantaneous ground state even when the initial condition is chosen to be the ground state of the initial Hamiltonian $H(0)$. Thus, at the end of the annealing schedule, when $A(1)=0$ and $B(1)>0$, the system is generally in an excited state with a number of kinks (defects) separating ferromagnetic domains (regions with aligned neighboring spins) when $J<0$, or kinks separating antiferromagnetic domains (regions with anti-aligned neighboring spins) when $J>0$.

\subsection{Kibble-Zurek mechanism and its extension}
\label{sec:KZM}
The Kibble-Zurek mechanism~\cite{Zurek2005} describes the process of kink formation assuming that the ratio of the parameters $A(s)$ and $B(s)$ changes linearly as a function of time near the critical point $A(s)/B(s)=1$. This assumption of linear time dependence in the vicinity of the critical point is reasonable because any analytical function can be expanded to linear order and we are interested in the system behavior near the critical point. Non-analytic driving schedules can also be accounted for within the KZM framework \cite{Diptiman08,Barankov08,Fernando19}.

Let us state the main theoretical predictions of the KZM and its generalization to be tested on the D-Wave device.
\begin{enumerate}
  \item 
    The kink density $\rho_{\rm kink}$, the number of kinks divided by the system size, follows the formula
    \begin{align}
        \rho_{\rm kink} \propto {t_a}^{-\frac{d\nu}{1+z\nu}}, 
        \label{eq:KZM1}
    \end{align}
    where $d$ is the spatial dimension, $\nu$ is the critical exponent for the correlation function, and $z$ is the dynamical critical exponent. In our case $d=\nu=z=1$, and thus:
    \begin{align}
        \rho_{\rm kink} \propto {t_a}^{-\frac{1}{2}}. 
        \label{eq:KZM2}
    \end{align}
 \item
   The $q$th cumulant $\kappa_q$ of the distribution function of the number of kinks $P(n)$, divided by the average $\kappa_1=\langle n\rangle $, is independent of the annealing time. In particular, for the one-dimensional transverse-field Ising model the second and third cumulants satisfy
   \begin{equation}
   \begin{split}
    \frac{\kappa_2}{\kappa_1} &=2-\sqrt{2}\approx 0.586,\\ \frac{\kappa_3}{\kappa_1}&=4-\frac{12}{\sqrt{2}}+\frac{8}{\sqrt{3}}\approx 0.134.
    \label{eq:cumulant_ratios}
   \end{split}
   \end{equation}
 \item
  Since the third and higher-order cumulants are small relative to the first and second order ones, the distribution function can be well approximated by a Gaussian distribution
  \begin{equation}
 P(n)=\frac{1}{\sqrt{2\pi (2-\sqrt{2})\langle n\rangle}}\exp\left[-\frac{(n-\langle n \rangle)^2}{2(2-\sqrt{2})\langle n \rangle}\right] .
 \label{eq:nd}
\end{equation}
\end{enumerate}

Below we briefly describe how these formulas are derived based on Refs.~\cite{Kibble1976,Zurek2005} for item 1, and on Ref.~\cite{Cui19} (its Supplementary Note 2 in particular) as well as on Ref.~\cite{DelCampo2018} for items 2 and 3, to provide pertinent physical background for our study. Readers interested only in the results can skip to Sec.~\ref{sec:exp-method}.

Second-order continuous phase transitions are characterized by the divergence of the correlation length $\xi$ and the  relaxation time $\tau$ at the critical point. Specifically,  as a function of the difference between the value of the control parameter $\lambda$ and its critical value $\lambda_c$, both quantities $\xi$  and $\tau$ exhibit a power-law behavior
\begin{align}
\xi=\xi_0|\varepsilon|^{-\nu},\quad \tau=\tau_0|\varepsilon|^{-z\nu},
\end{align}
where $\varepsilon=(\lambda-\lambda_c)/\lambda_c$, and $\xi_0$ and $\tau_0$ are constants. The divergence of the relaxation time introduces a separation of time scales and allows one to describe the crossing of the phase transition as a sequence of stages:
In the first stage, far from criticality where $|\epsilon|$ is not very small, the relaxation time is not large and system follows the instantaneous equilibrium state, the ground state in the context of quantum phase transition at zero temperature. The system evolves adiabatically.  Then, in the second stage, as $|\epsilon|$ becomes small, the relaxation time grows rapidly and the state of the system has no time to relax to the ground state, and the system becomes effectively frozen.  As the parameter further changes, the system enters the final third stage, and $|\epsilon|$ again becomes large, thus the dynamics becomes adiabatic again.  This is the so-called adiabatic-impulse approximation  \cite{Zurek2005,Damski2006}.

The key testable prediction of the KZM is that, after crossing the phase transition in the second stage, the average length scale in which the order-parameter is uniform is set by the equilibrium value of the correlation length when the system unfreezes at the point where the third stage is reached.

To formulate this idea quantitatively, consider a driving scheme such that the distance to the critical point varies linearly in time according to $\varepsilon=(t-t_c)/t_a$ on a timescale $t_a$, where $t_c$ denotes the instant when the system parameters cross the critical point.
Equating the instantaneous equilibrium relaxation time $\tau (\hat{t})$ to $\hat{t}-t_c$, the time elapsed after crossing the critical point, yields the freeze-out time scale $\hat{t}-t_c=(\tau_0t_a^{z\nu})^{1/(1+z\nu)}$, which yields the average correlation length as $\hat{\xi}\equiv \xi(\hat{t}) =\xi_0(\tau_0/t_a)^{\nu/(1+z\nu)}$. 
A kink may form at the interface between different domains of  size $\hat{\xi}$.
Then the kink density is given by the inverse of the volume $\hat{\xi}^d$ and scales as a universal power-law \cite{Kibble1976,Zurek1985}
\begin{align}
\label{eq:KZM_original}
\rho_{\rm kink}=\frac{1}{\hat{\xi}^d}\propto\left(\frac{\tau_0}{t_a}\right)^{\frac{d\nu}{1+z\nu}},
\end{align}
for a system in $d$ spatial dimensions. This is Eq. (\ref{eq:KZM1}). When the system size is $L^d$, the average number of kinks is thus $\langle n\rangle=\rho_{\rm kink}L^d$. 

This picture applies both to classical and quantum phase transitions. In the quantum case,  the relaxation time is identified with the inverse of the energy gap between the ground state and the first excited state that closes at the critical point, and the KZM describes the critical dynamics as well~\cite{Polkovnikov05,Damski05,Dziarmaga05,Zurek2005}.

The above physical picture is quite generic and the result is  valid independent of the details of the system Hamiltonian. If we restrict ourselves to the quantum phase transition of the one-dimensional transverse-field Ising model, more detailed information can be extracted on the distribution of kink numbers as follows \cite{DelCampo2018,Cui19}.

The one-dimensional transverse-field Ising model with a periodic boundary can be solved (diagonalized) under periodic boundary condition by the Jordan-Wigner transformation \cite{Nishimoribook2011}, which rewrites the spin operators in terms of spinless fermion operators.  Kinks appear in pairs under periodic boundary, and we therefore consider the number of kink pairs, which is described by the operator (for the ferromagnetic case)
\begin{align}
    \mathcal{N}=\frac{1}{4}\sum_{j=1}^{L}\big(1-\sigma_j^z \sigma_{j+1}^z\big)= \sum_{k\ge 0}\gamma_k^{^\dagger} \gamma_k,
\end{align}
where $L$ is the total number of sites and $\gamma_k^{\dagger}$ and $\gamma_k$ are creation and annihilation operators of fermions. The distribution function of kink pairs is defined by
\begin{align}
    P(n)={\rm Tr}\big( \rho \,\delta (\mathcal{N}-n) \big),
\end{align}
where $\rho$ is the density matrix for the state after annealing. It helps to use the characteristic function $\tilde{P}(\theta)$, the Fourier transform,
\begin{align}
    P(n)=\frac{1}{2\pi}
    \int_{-\pi}^{\pi} d\theta \, \tilde{P}(\theta)\, e^{-in\theta}.
\end{align}
Since kink pairs with different wave numbers are independent, the characteristic function is decomposed into a product
\begin{align}
    \tilde{P}(\theta)=\prod_{k\ge 0}\left[1+(e^{i\theta}-1)p_k\right],
    \label{eq:cf1}
\end{align}
where
\begin{align}
    p_k=\langle \gamma_k^{\dagger}\gamma_k \rangle =e^{-2\pi Jt_ak^2/\hbar}.
\end{align}
We have used the Landau-Zener formula for the creation of a kink pair. Equation (\ref{eq:cf1}) indicates that the number of kink pairs follows the Poisson binomial distribution.  Then the cumulants are easily evaluated, the first three of which are
\bes
\begin{align}
    \tilde{\kappa}_1&=\langle n\rangle =\sum_{k\ge 0}p_k, \\
    \tilde{\kappa}_2& =\sum_{k\ge 0}p_k (1-p_k),\\
    \tilde{\kappa}_3&=\sum_{k\ge 0}p_k (1-p_k)(1-2p_k).
\end{align}
\ees
In the long time scale limit $t_a \gg \hbar/(2\pi^3 J)$, the first cumulant (the average), reduces to
\begin{align}
    \tilde{\kappa}_1&=\sum_{k \ge 0} p_k =\frac{L}{2\pi} \int_0^{\pi}dk\, e^{-2\pi Jt_ak^2/\hbar} \nonumber\\
     & \to \frac{L}{2\pi}\int_0^{\infty} dk\, e^{-2\pi Jt_ak^2/\hbar}=\frac{L}{4\pi} \sqrt{\frac{\hbar}{2Jt_a}}.
\end{align}
This is consistent with Eq.~\eqref{eq:KZM2} of the KZM.  Similarly, the second and third cumulants are evaluated to yield
\bes
\begin{align}
    \tilde{\kappa}_2&= \left(1-\frac{1}{\sqrt{2}}\right) \tilde{\kappa}_1,\\
        \tilde{\kappa}_3&= \left(1-\frac{3}{\sqrt{2}}+\frac{2}{\sqrt{3}}\right) \tilde{\kappa}_1.
\end{align}
\ees
The cumulants $\kappa_q$ for the number of kinks can be derived from the above cumulants for the number of kink pairs as $\kappa_q=2^q \tilde{\kappa}_q$,
\bes
\begin{align}
    \kappa_1 &= 2\tilde{\kappa}_1 =\frac{L}{2\pi} \sqrt{\frac{\hbar}{2Jt_a}},\\
    \kappa_2&=4\tilde{\kappa}_2 =(2-\sqrt{2})\kappa_1,\\
    \kappa_3&=8\tilde{\kappa}_3=\left(4-\frac{12}{\sqrt{2}}+\frac{8}{\sqrt{3}}\right) \kappa_1.
\end{align}
\ees
These give Eq.~\eqref{eq:cumulant_ratios}.

The Gaussian distribution Eq.~\eqref{eq:nd} follows from  setting to zero all cumulants $\kappa_q$ with $q\ge 3$, a reasonable approximation as their value is much smaller than $\kappa_1$ and $\kappa_2$; see \cite{Cincio07,DelCampo2018,Cui19}.

\subsection{Experimental methods}
\label{sec:exp-method}

We used two different DW2KQ devices, one located at the NASA Ames Research Center and the other at D-Wave Systems, Inc. in Burnaby. The latter is a lower-noise version of the former (for documentation see Ref.~\cite{DW-2000Q-5}).
The D-Wave Chimera graph comprises $\ell\times \ell$ unit cells of sparsely connected $K_{4,4}$ bipartite graphs, for a total of $8\ell^2$ qubits, each coupled to up to $6$ other qubits. We chose four chain lengths: $L=50$, $200$, $500$, and $800$. For each  size we generated $200$ instances of configurations of the one-dimensional chain with a free boundary by self-avoiding random walks starting from a randomly selected qubit on the Chimera graph. For each of these $200$ instances, we carried out $1,000$ annealing cycles at a given annealing time $t_a$. Thus, we generated $200,000$ samples for each $t_a$ and $L$, and recorded the distribution (histogram) of the kink density. The annealing time $t_a$ ranges from $1\mu$s to $2$ms, for a total of $33$ values. 

We tested three cases of the coupling parameter $J$: ferromagnetic ($J=-1$), antiferromagnetic ($J=1$), and randomly chosen gauges. The latter starts from ferromagnetic interactions, then half of the qubits are chosen randomly and the signs of their interactions are flipped. This prescription is meant to cancel (unintended, device-specific) local biases toward a specific direction at each qubit. As shown in Appendix~\ref{sec:Appendix1},  the antiferromagnetic and random-gauge cases give almost identical results, while the ferromagnetic case tends to exhibit unstable behavior. We therefore show results for the antiferromagnetic case in the main text.

\section{Average kink density}\label{sec:pl}

\begin{figure*}[t]
\includegraphics[width=\textwidth]{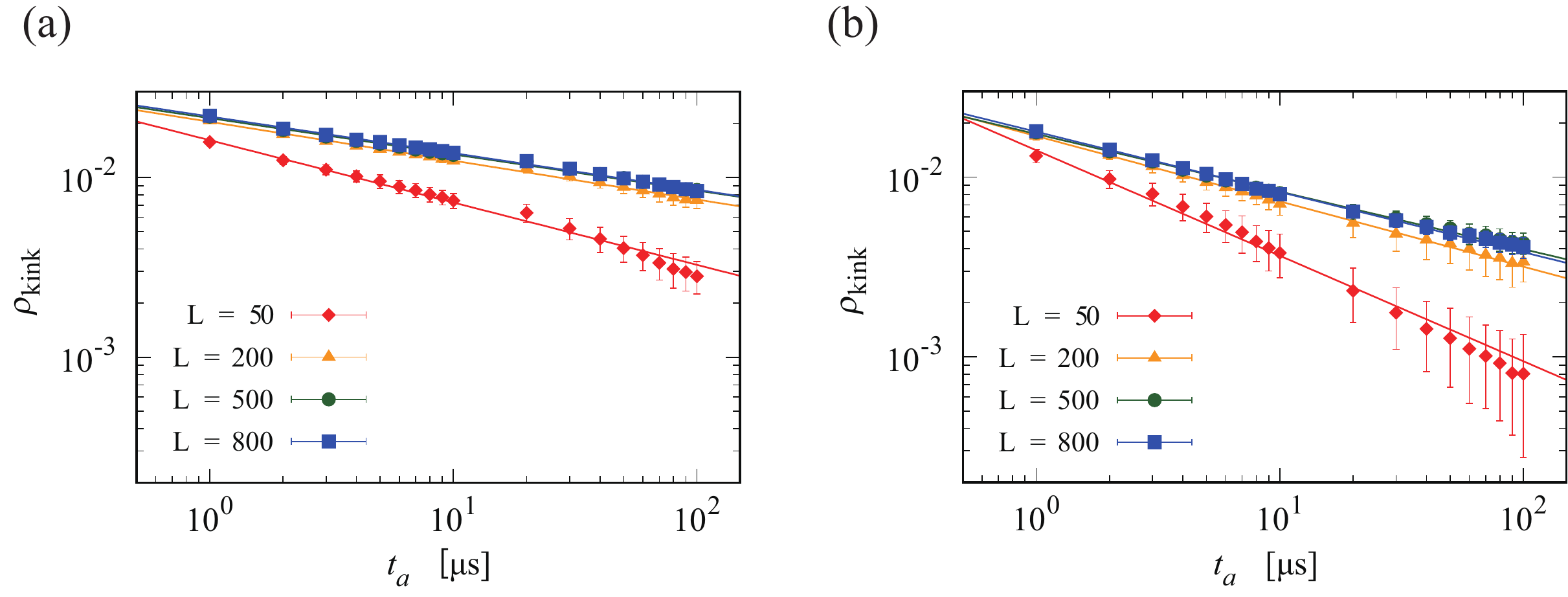}
\caption{Kink density as a function of annealing time (log-log scale). Error bars indicate the 68\% confidence interval. (a) Data from the~\texttt{C16} solver on the DW2KQ at NASA. (b) Data from the~\texttt{D\_2000Q\_5} solver on the DW2KQ in Burnaby. Data is averaged over $200,000$ samples at each value of $t_a$.}
\label{fig:pl}
\end{figure*}

The average kink density as a function of the annealing time $t_a$ and for different sizes $L$ is shown in Fig.~\ref{fig:pl}(a) for the NASA device and Fig.~\ref{fig:pl}(b) for the Burnaby device.  We analyze the data for the time range $t_a \le 100~\mathrm{\mu}$s because the data beyond 100~$\mathrm{\mu}$s show different, less stable, behavior.  Likely reasons include the effect of $1/f$ noise, which becomes apparent at long annealing times, and a significant increase in the persistent current for $t_a > 100~\mathrm{\mu}$s~\cite{DW-2000Q-5}, which reduces qubit coherence.  See Appendix~\ref{sec:Appendix1} for data beyond $100~\mathrm{\mu}$s and a more detailed discussion.

The KZM assumes that the number of kinks is at least $1$ on average, which means that the inequality $\rho_{\rm kink}>1/L$ should hold. We therefore also exclude the data for $L=50$ from the analysis because the kink density is too low: we find empirically (see Fig.~\ref{fig:pl}) that $\rho_{\rm kink}<1/L=0.02$, which implies that the KZM does not apply.

A first qualitative observation from Fig.~\ref{fig:pl} is that the kink density obeys a power law. It is also clearly seen that the kink density is lower on the Burnaby device in Fig.~\ref{fig:pl}(b) than on the NASA device in Fig.~\ref{fig:pl}(a) for the same parameter values $L$ and $t_a$.  This is in accordance with the `low-noise' characteristics of the Burnaby device~\cite{DW-2000Q-5}.

Delving deeper into quantitative aspects, we fit the data to the power law 
\begin{align}
    \rho_{\rm kink}\propto {t_a}^{-\alpha}
    \label{eq:power_law}
\end{align} and evaluate the exponent $\alpha$. The result is given in Table~\ref{tab:pl}.

\begin{table}[t]
\caption{Results from D-Wave device runs for the exponent $\alpha$ of the power-law scaling describing the decay rate of the kink density as shown in Fig.~\ref{fig:pl}. }
\begin{ruledtabular}
\begin{tabular}{ccc}
$L$  &  NASA  & Burnaby \\
\hline
50  & 0.347$\pm$0.008 &  0.587$\pm$0.016 \\
200 & 0.216$\pm$0.003 & 0.363$\pm$0.003 \\
500 & 0.201$\pm$0.003 & 0.320$\pm$0.005 \\
800 & 0.204$\pm$0.002 & 0.335$\pm$0.003 \\
\end{tabular}
\end{ruledtabular}
\label{tab:pl} 
\end{table}
Apart from the case of $L=50$, the exponent $\alpha$ is almost independent of the chain length $L$. The NASA device has $\alpha\approx 0.20$ ($L=800$) and the Burnaby device has $\alpha\approx 0.34$  ($L=800$). These values are far from the KZM prediction of $0.5$ in Eq.~\eqref{eq:KZM2}. Preliminary numerical simulations under unitary dynamics suggest that the value of  $\alpha$ may not be attributed to the nonlinear functional form of $A(s)$ and $B(s)$. As the schedules can be effectively linearized, corrections to KZM behavior resulting from nonlinear passage across the critical point \cite{Diptiman08,Barankov08} may be  ruled out. It is thus reasonable to suspect that the difference originates from deviations from unitary dynamics that are not accounted for in the theory. 

A natural first step is therefore to incorporate the coupling of qubits to the environment, for which we use the standard spin-boson model with the following Hamiltonian~\cite{Leggett1987}:
\begin{align}
    H_{\rm total}=H(s)+\sum_{i,k}\big\{C_k(a_{i,k}^{\dagger}+a_{i,k})\sigma_i^z +\hbar\omega_{i,k}a_{i,k}^{\dagger}a_{i,k}\big\},
 \label{eq:Hamiltonian_coupling_boson}
\end{align}
where $H(s)$ is the original Hamiltonian of Eq.~\eqref{eq:original_Hamiltonian}. 
Independent bosons (harmonic oscillators) with frequency $\omega_{i,k}$ couple to the $z$ component of the $i$th Pauli matrix. The coefficient $C_k$ is assumed to have an Ohmic spectrum,
\begin{align}
    J(\omega)=\frac{4\pi}{\hbar^2} \sum_i C_k^2 \delta (\omega-\omega_k)=2\pi \eta \omega\quad (\omega <\omega_{\rm c})
  \label{eq:Ohmic}
\end{align}
with the sharp cutoff frequency $\omega_{\rm c}$ and the coupling constant $\eta$. 

The Ohmic spin-boson approach has been successfully used many times in modeling the dynamics of open system quantum annealing~\cite{lanting2011,ABLZ:12-SI,q-sig,q-sig2,Albash:2014if,Albash:2015pd,Amin:2015qf,Albash:2015nx,Boixo2016,Munoz-Bauza:2019aa}. In particular, Ref.~\cite{Mishra2018} reported a closely related open quantum systems study of transverse field Ising spin chains  with alternating sectors of strong/weak ferromagnetic coupling, but this study did not include a comparison to KZM theory.

Ground-state (time-independent) properties of the above model have already been studied by quantum Monte Carlo simulations~\cite{Werner2005} and renormalization group methods~\cite{Pankov2004,Sachdev2004}. The conclusion of these papers is that the quantum phase transition persists under a bosonic environment and the values of the critical exponents change from $\nu=z=1$ for the isolated system to $\nu=0.64$ and $z=1.99$ for the system coupled to a \textit{zero temperature} bosonic environment. Note the sharp contrast with other models of decoherence which do not alter the bare critical exponents and lead to environmentally-induced heating~\cite{Patane08,Nalbach15,Dutta16}. 

\begin{figure*}[t]
\includegraphics[width=\textwidth]{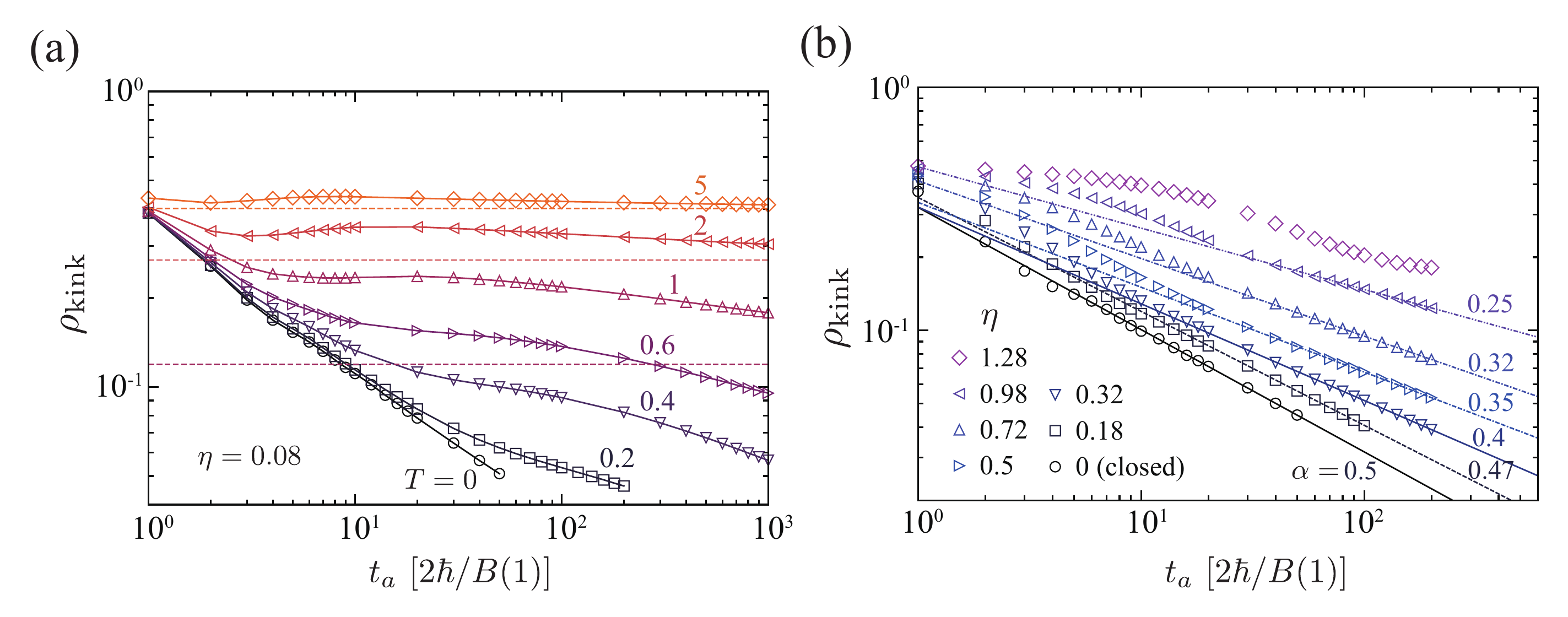}
\caption{\label{fig:iTEBD} Numerically computed kink density as a function of annealing time (log-log scale), using iTEBD with QUAPI. A linear annealing schedule, $A_{\rm lin}(s)/2 = 1 - s$ and $B_{\rm lin}(s)/2 = s$, is employed here, where the unit of energy is given by $B(1)/2$ with $B(1)$ provided in Fig.~\ref{fig:schedule}. The Ohmic cutoff frequency is $\omega_{\rm c} = 5$ [$B(1)/2\hbar$], the Trotter time slice is $\Delta t = 0.05$ [$2\hbar/B(1)$], the cutoff memory time is $\tau_{\rm c} = 10$ [$2\hbar/B(1)$], and the bond dimension is up to $128$. (a) Results for various temperatures and a fixed coupling strength $\eta = 0.08$.  Dashed horizontal lines show the thermal expectation values at temperatures $T = 5$, $2$, and $1$ from the top. The unit of temperature is given by $B(1)/2 k_B$. (b) Results for  zero temperature and various coupling strengths. The rightmost eight data points for each coupling strength are fitted with the power law, Eq.~\eqref{eq:power_law}, and the corresponding exponent $\alpha$ is provided above each data set shown. }
\end{figure*}

\begin{figure}[t]
\includegraphics[width=.4\textwidth]{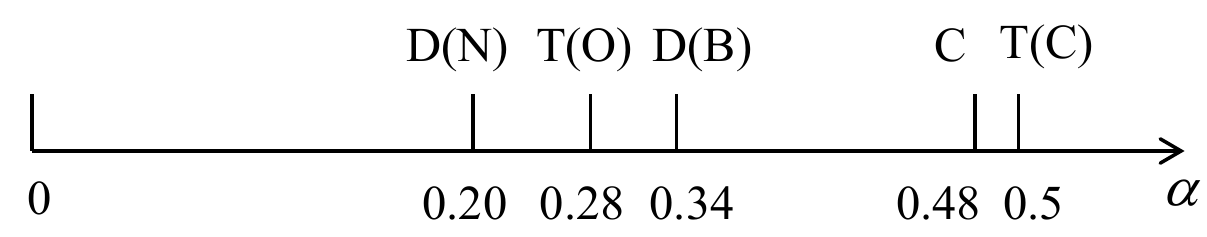}
\caption{\label{fig:exponents} The values of the decay exponent $\alpha$ by various methods. D(N): Data from DW2K at NASA. T(O): Open-system theory. D(B): Data from DW2K at Burnaby.  C: Classical SVMC (Sec. \ref{sec:SVMC}). T(C): Closed-system theory.}
\end{figure}

The modified values  of the critical exponents $\nu$ and $z$ are independent of the coupling constant $\eta (>0)$ in Eq.~\eqref{eq:Ohmic}.  We assume that the KZM applies to the present open system case because KZM theory is developed based only on the divergence of the relaxation time near a critical point, without recourse to a microscopic Hamiltonian. We therefore apply the generic Eq.~\eqref{eq:KZM_original} to find the exponent $\nu/(1+z\nu)=0.28$, about half of the isolated case of $0.5$ in Eq.~\eqref{eq:KZM2}.
Although this open-system theoretical value of $0.28$ is still different from the experimental values of $0.20$ (NASA) and $0.34$ (Burnaby), the spin-boson model of Eq.~\eqref{eq:Hamiltonian_coupling_boson} significantly reduces the difference between theory and experiment, in comparison with the closed-system theoretical value of $0.5$ as illustrated in Fig.~\ref{fig:exponents}.
It is therefore reasonable to conclude that the Hamiltonian Eq.~\eqref{eq:Hamiltonian_coupling_boson} captures, to a first approximation, the essential features of the behavior of the D-Wave devices near the critical point of the one-dimensional transverse-field Ising model embedded on the Chimera graph. 
To achieve more precise quantitative agreement between theory and experiment, we would need to incorporate additional elements that have not been taken into account so far.  Such may include (i) finite temperature effects not considered in Refs.~\cite{Werner2005,Pankov2004,Sachdev2004}; (ii) transient phenomena due to the finite annealing time; and (iii) control errors, i.e., imprecision in the parameter setting in the devices~\cite{Pearson:2019aa}.

The impacts of the first two items (i) and (ii) listed above were studied by some of us in Refs.~\cite{Suzuki2019,Oshiyama}.
Extensive numerical computations using the time-evolving block decimation (TEBD) method, as well as infinite-TEBD (iTEBD) combined with the quasi-adiabatic propagator path integral (QUAPI) reveal that, as shown in Fig. \ref{fig:iTEBD}, (a) the kink density approaches a temperature-dependent constant as $t_a$ becomes very large; (b) the kink density may behave non-monotonically as a function of $t_a$ in the transient time range if the temperature is finite; (c) the effective exponent $\alpha$ in $\rho_{\rm kink}\propto {t_a}^{-\alpha}$ around a given time $t_a$ depends on the coupling strength even when the temperature is zero.

More precisely, Fig.~\ref{fig:iTEBD}(a) shows the temperature dependence of $\rho_{\rm kink}$ for a fixed coupling strength $\eta$ obtained by iTEBD with QUAPI. One can see that the curve of $\rho_{\rm kink}$ for finite temperature deviates upwards from that of the zero temperature case with increasing $t_a$ and the deviation is more pronounced for higher temperatures. The results for the temperatures $T = 1$, $2$, and $5$ [in units of $B(1)/2 k_B$] imply that $\rho_{\rm kink}$ behaves non-monotonically with $t_a$ and would approach the thermal average, $[1 - \tanh (B(1)/2k_BT)]/2$, as $t_a\to\infty$. Since our data in Fig.~\ref{fig:pl} do not show an approach to a constant, we may conclude that temperature effects in the form considered in Ref.~\cite{Suzuki2019} have not come into play in our data for the present range of annealing time.\footnote{See Sec.~\ref{section:Effective_temperature} for the evaluation of the effective temperature in a different sense.}

Regarding the observation (c), Fig.~\ref{fig:iTEBD}(b) shows the dependence of the slope $\alpha$ on the coupling strength $\eta$ at zero temperature. Note the transient effects, which increase in magnitude with $\eta$, and also extend to larger $t_a$. We would expect the exponent $\alpha$ to approach a constant independent of the coupling constant for sufficiently large $t_a$, if we assume consistency with the equilibrium computations in Refs.~\cite{Werner2005,Sachdev2004,Pankov2004}, which suggest a universal exponent $\alpha=0.28$ independent of $\eta$ as mentioned above.  We suspect that the deviations of our experimental result $0.20$ and $0.34$ for the exponent $\alpha$ from the theoretical equilibrium value of $0.28$ are at least in part a result of transient effects. These effects are difficult to analyze in a precise way because the effective exponent changes as a function of the coupling constant $\eta$ and the annealing time range, and is therefore non-universal as seen in Fig.~\ref{fig:iTEBD}(b) and Fig.~1 of Ref.~\cite{Suzuki2019}.

Noise amplitude and control errors may qualitatively explain the difference between the NASA and Burnaby devices. The latter is a newer, low-noise model, with lower $1/f$ noise amplitude and more accurate control~\cite{DW-2000Q-5}. Better control in the specification of system parameters, the interaction strength $J$ between neighboring qubits as well as the local longitudinal field (which is nominally zero in the present problem), results in less randomness in the problem parameters, which should lead to a more rapid decrease of the kink density as a function of annealing time on the Burnaby device, meaning a larger exponent $\alpha$. Moreover, the less noisy Burnaby device has a value closer to that of the closed-system value, and so it stands to reason that the fact that the spin-boson value is intermediate between the noisier NASA device and the Burnaby device is a reflection of the fact that the former device is more closely described as an open system than the latter. The extracted exponents of $0.34$ (Burnaby) and $0.20$ (NASA) are consistent with this picture.

Note that randomness in $J$ from location to location necessarily induces more kinks and eventually leads to a very slow inverse-logarithmic law, instead of a polynomial decay~\cite{Huse1986,Dziarmaga2006,Caneva07}.

\section{Kink distribution}
\label{sec:dist}

We collected  statistics of the kink density from $200,000$ samples for each $t_a$ at $L=800$ as a test of the distribution function theory developed by one of us as described in Sec.~\ref{sec:KZM} and Ref.~\cite{DelCampo2018}. See also Ref.~\cite{Cui19}.  One of the important predictions of these references is that the ratio of the $q$th cumulant $\kappa_q$ ($q\ge 2$) to the first cumulant $\kappa_1$, the average, is independent of the annealing time $t_a$ [Eq.~\eqref{eq:cumulant_ratios}]. 

\begin{figure*}[t]
\includegraphics[width=\textwidth]{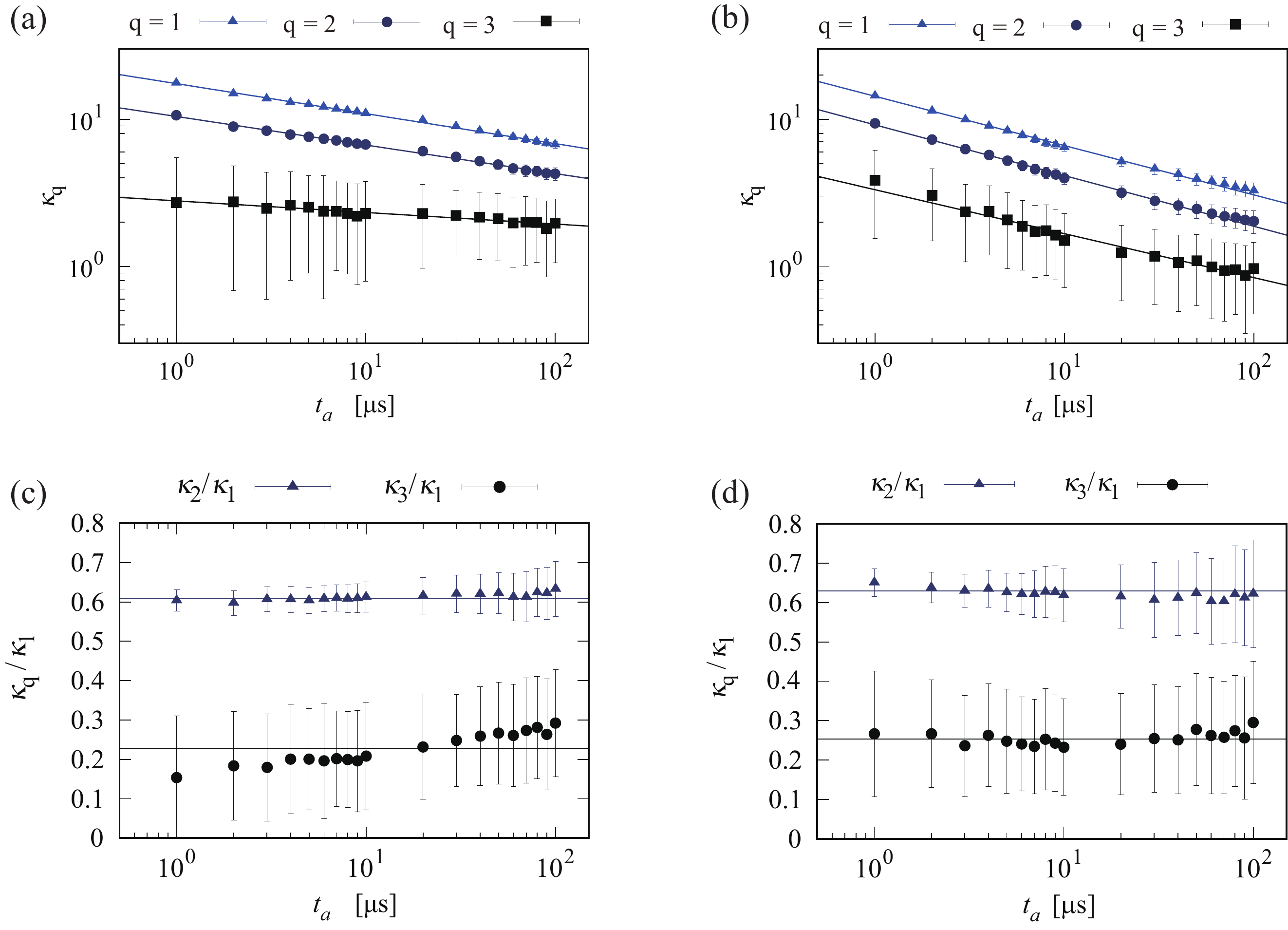}
\caption{Cumulants of $q$th order $\kappa_{q}$ of the kink distribution. The chain length is $L=800$. Error bars indicate the $68$\% confidence interval. Power-law scaling of cumulants from the first-order $\kappa_{1}$ to third-order $\kappa_{3}$ as functions of the annealing time $t_a$ on (a) the DW2KQ at NASA and (b) the DW2KQ in Burnaby. Ratios $\kappa_{2}/\kappa_1$ and $\kappa_{3}/\kappa_1$ of cumulants on (c) the DW2KQ at NASA and (d) the DW2KQ in Burnaby. The solid lines are optimized fits to constants, $\kappa_2/\kappa_1\approx0.61$ and $\kappa_3/\kappa_1\approx0.23$ for the NASA case, and $\kappa_2/\kappa_1\approx0.63$ and $\kappa_3/\kappa_1\approx0.25$ for the Burnaby case. }
\label{fig:cumulants} 
\end{figure*}

Figure~\ref{fig:cumulants} shows the $t_a$ dependence of three cumulants [panels (a) and (b) for the NASA and Burnaby devices, respectively] and the ratios $\kappa_2/\kappa_1$ and $\kappa_3/\kappa_1$ [panels (c) and (d) for the NASA and Burnaby devices, respectively]. With the exception of $\kappa_3/\kappa_1$ for the NASA device, these ratios indeed appear to be independent of $t_a$, as predicted. The experimental values are extracted as $\kappa_2/\kappa_1\approx 0.61-0.63$ and $\kappa_3/\kappa_1\approx 0.23-0.25$. The theoretical predictions are $0.586$ and $0.134$, respectively, so the former ratio is closer to the theoretical prediction than the latter.  A possible reason is the large uncertainty in statistics as reflected in the large error bars in Figure~\ref{fig:cumulants}(c) and (d) for $\kappa_3/\kappa_1$. Indeed, the lower ends of the error bars of this ratio lie around $0.1$, and the theoretically predicted value of $0.134$ is within the error bars.  Apart from this subtlety, the experimental data are consistent with the theory presented in Ref.~\cite{DelCampo2018} (see also Refs.~\cite{Cui19,GomezRuiz19b}).

\begin{figure}[t]
\includegraphics[width=\columnwidth]{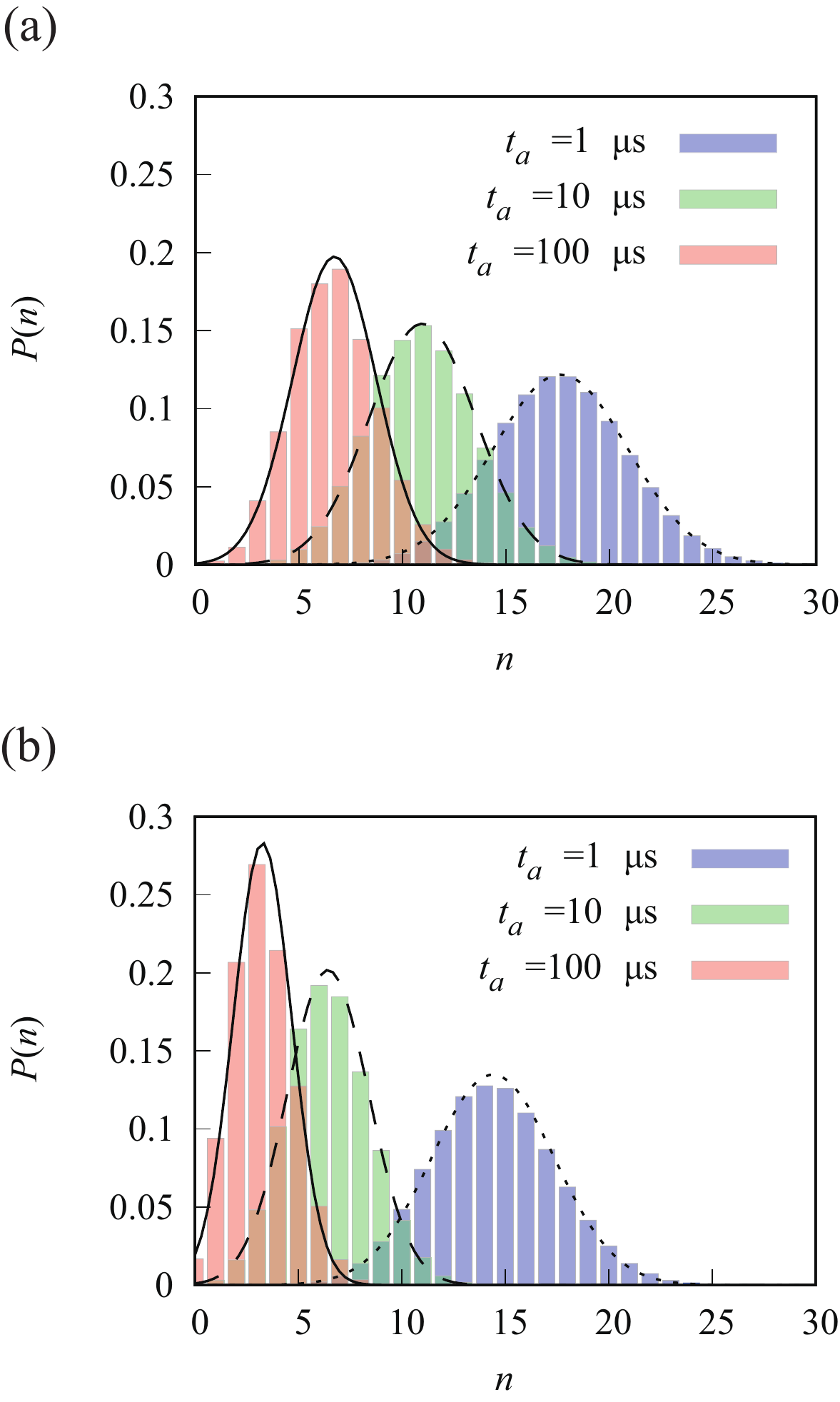}
\caption{Histograms of the number of kinks observed on (a) the DW2KQ at NASA and (b) the DW2KQ in Burnaby. The chain length is $L=800$. From right to left, the annealing times $t_a$ are $1,10,100~\mu $s, respectively. Solid, dashed, and dotted lines are the Gaussian distributions of Eq.~\eqref{eq:nd}.
}
\label{fig:dist} 
\end{figure}

Since the third and higher order cumulants are much smaller than the second order cumulant, the distribution is well approximated by the Gaussian distribution function~\eqref{eq:nd}.
Figure~\ref{fig:dist} shows the distributions at three values of $t_a$. All three cases are very well approximated by this Gaussian, as drawn in solid, dashed, and dotted lines.
\footnote{The distribution is close to Gaussian but is not exactly so due to the small but non-vanishing value of the third cumulant.}
Additional data are presented in Appendix~\ref{sec:Appendix2}.

\begin{figure}[t]
\includegraphics[width=\columnwidth]{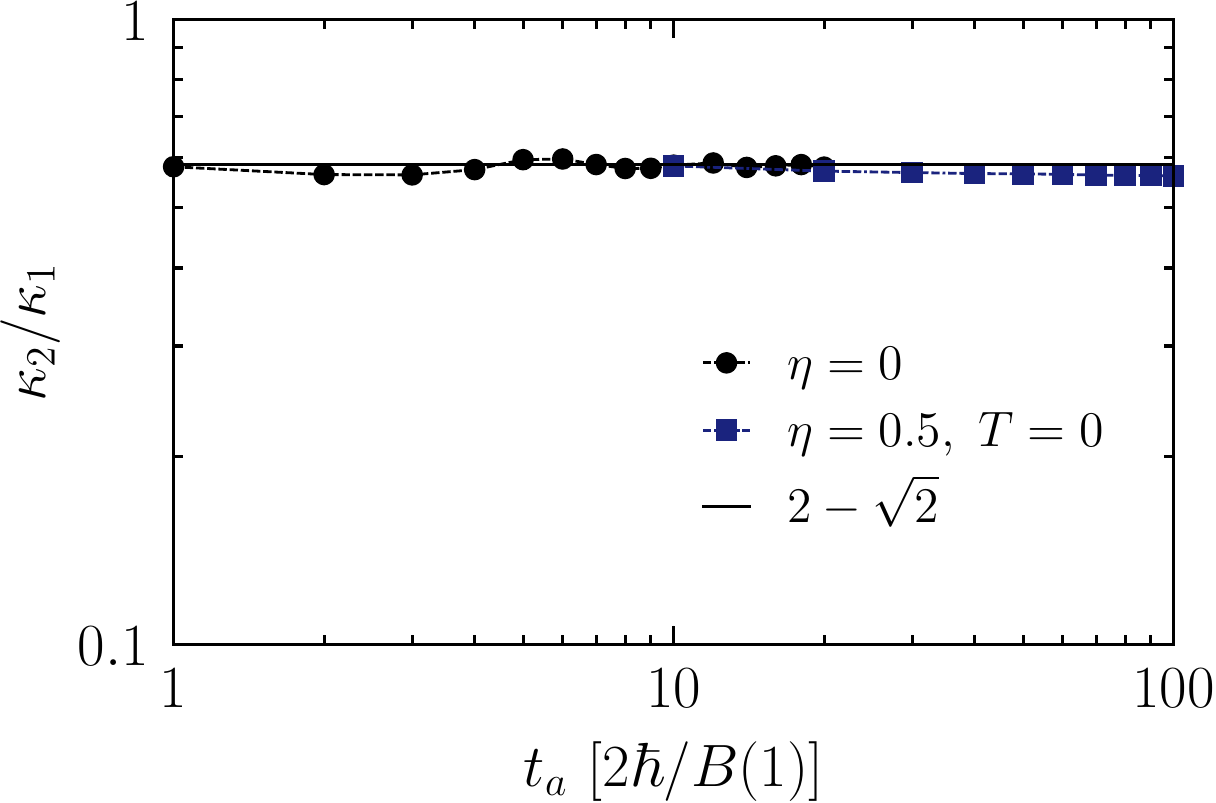}
\caption{The cumulant ratio $\kappa_2/\kappa_1$ as a function of the annealing time computed by iTEBD with QUAPI. The annealing schedule and other parameters are the same as in Fig.~\ref{fig:iTEBD}. $\eta$ is the strength of the spin-boson coupling. The solid horizontal line denotes $\kappa_2/\kappa_1=2-\sqrt{2}$, which is the value theoretically predicted for an isolated (closed) system.
}
\label{fig:dist_iTEBD} 
\end{figure}

It is remarkable that we find such strong agreement between the closed-system quantum theory of Ref.~\cite{DelCampo2018} and the experimental results for the kink decay, cumulants, and distribution, given that the experiment is conducted on devices whose behavior is described by open-system dynamics as discussed in the previous section.
This suggests that these features are robust aspects of the kink statistics that lie beyond the KZM theory. 
This is the first time that a quantum simulator predicted a hitherto unknown phenomenon.
To confirm the reliability of this result, we have conducted extensive numerical computations using iTEBD with QUAPI. The result is shown in Fig.~\ref{fig:dist_iTEBD}.
It is clearly seen that the ratio $\kappa_2/\kappa_1$ is constant as a function of the annealing time and the constant value is independent of whether or not the system is coupled to the environment. Although it is difficult to compute cumulants beyond second order due to the large number of terms that must be summed, the present result supports the experimental finding  that the theoretical prediction in Ref.~\cite{DelCampo2018} holds beyond its scope of an isolated system, at least for $\kappa_2/\kappa_1$.

\section{Tests of classicality}
\label{section:classical}

In this section we address whether our empirical results and reasonable agreement with a quantum theory of the KZM can also be understood using a purely classical approach. We first consider a Boltzmann distribution of the kink density of the classical Ising chain, then the classical spin-vector Monte Carlo model~\cite{SSSV}, which has been successfully applied to at least partially describe the outcomes of experiments using the D-Wave devices in past studies~\cite{q-sig2,Albash:2014if,Boixo2016,Albash:2017aa,Mishra2018,DW2000Q,King:2016aa}, and also in recent theoretical studies of quantum annealing~\cite{Susa2018,Yamashiro:2019aa}.

\subsection{Boltzmann distribution and effective temperature of the kink distribution}
\label{section:Effective_temperature}

A question of significant interest, e.g., due to the potential for quantum-assisted classical machine learning applications of quantum annealers as Boltzmann machines~\cite{Amin:2016,Benedetti:2016oz,Li:2019}, is whether the kink distribution is thermal and well described by a Boltzmann distribution. Various previous studies have found mixed results in terms of trying to fit such thermal distributions to empirical quantum annealing data, an issue that is understood in terms of the fact that the distribution freezes once quantum fluctuations cease at low (but non-zero) values of the transverse field~\cite{Amin:2015qf,Marshall:2017aa,Li:2019,Vinci:2017ab}. The associated effective temperature is a relevant metric for quantifying how noisy one quantum annealer is relative to another. Given the result we found in Sec.~\ref{sec:pl}, that the Burnaby device generates fewer kinks than the NASA device, we expect the former to exhibit a lower effective temperature. In this section we address these issues, and provide an assessment of how well a trivial classical model matches the empirical kink distribution we have observed.

Let the empirical distribution be denoted by $P(n;t_a)$, where $n$ is the number of kinks, and let  $Q(n;\beta')$ denote the Boltzmann distribution for the classical Ising model of a chain of length $L$. The latter is easily shown to be:
\begin{equation}
Q\left(n;\beta'\right)=g(n)\frac{e^{-\beta'E(n)}}{Z},
\label{eq:Gibbs}
\end{equation}
where $g(n) = \binom{L-1}{n}$ is the degeneracy of $n$ kinks, $Z=( e^{\beta'}+e^{-\beta'})^{L-1}$ is the partition function, $E(n)=2n+1-L$ is the dimensionless energy, and $\beta'$ is:
\begin{equation}
\beta'=\frac{B(1)}{2}\frac{1}{k_B T}.
\end{equation}
We write $\beta'$ instead of $\beta$ to indicate that this is a dimensionless effective inverse temperature reflecting all noise effects, not the inverse of the real physical temperature. $B(1)$ is the device-dependent value of the Ising schedule [Eq.~\eqref{eq:original_Hamiltonian}] at the end of the anneal (at $s=1$), $k_B$ is the Boltzmann constant, and $T$ is the physical temperature.  $T=12.1$~mK and $B(1)/2 = 6.344$~GHz for the NASA DW2KQ device; $T=13.5$~mK and  $B(1)/2 = 5.930$~GHz for the DW2KQ in Burnaby.

To estimate $\beta'$, we minimize (with respect to $\beta'$) the Kullback-Leibler (KL) divergence and the trace-norm distance between the experimental distribution and the Boltzmann distribution in Eq.~\eqref{eq:Gibbs}. The KL divergence $D_{\rm KL}$ and the trace-norm $D_{\rm TN}$ are defined by using the empirical distribution $P(n;t_a)$ and $Q\left(n,\beta'\right)$ as follows:
\bes
\begin{eqnarray}
D_{\rm KL}(t_a)&=&\sum_{n} P(n;t_a)\log\frac{P(n;t_a)}{Q\left(n;\beta'\right)},\label{eq:KL}\\
D_{\rm TN}(t_a)&=&\frac{1}{2}\sum_{n}\left| P(n;t_a)-Q\left(n;\beta'\right)\right|.\label{eq:TN}
\end{eqnarray}
\ees
To obtain reliable estimates of the effective temperature, we first minimize the KL divergence to obtain the first approximation of the effective temperature $1/\beta'$, because the KL divergence turns out to be robust against data fluctuations. The $\beta'$ thus obtained is then used as the initial value in the effective temperature estimation based on the trace norm.

\begin{figure}[t]
\includegraphics[width=\columnwidth]{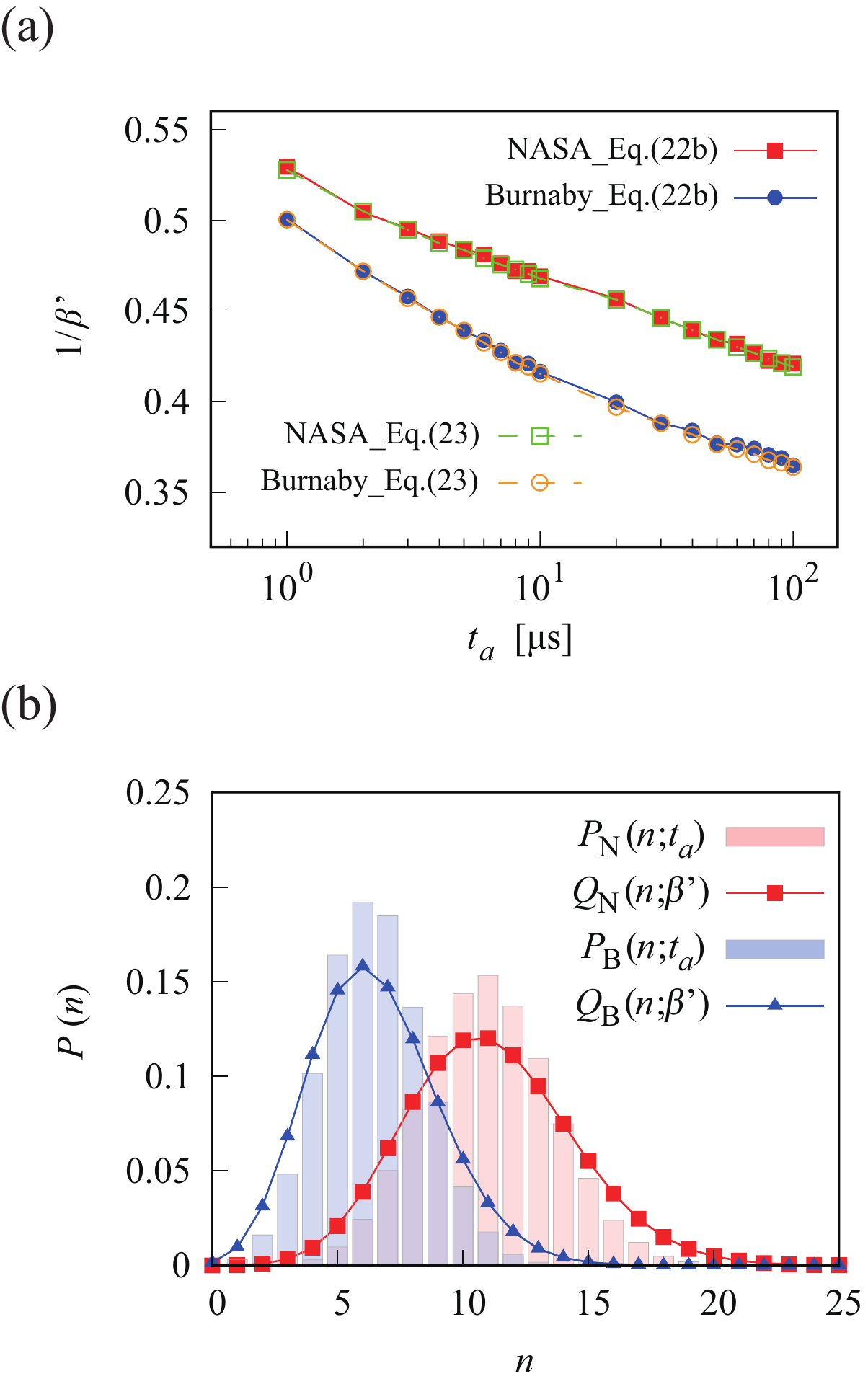}
\caption{\label{fig:effT}Effective temperatures and the optimized Boltzmann distributions on the D-Wave device. The chain length is $L=800$. (a) Effective temperature $1/\beta'$ as a function of the annealing time $t_a$. Effective temperatures are obtained by equating Eq.~(\ref{eq:rho_kink_Boltzmann}) to the D-Wave data or minimizing Eq.~(\ref{eq:TN}) with respect to $\beta'$. (b) The optimized Boltzmann distributions at $t_a=10~\mu$s and the best fit of the Boltzmann distribution. Here, $P_{\rm N}$ and $P_{\rm B}$ are observed distributions in the DW2KQ at NASA and Burnaby, respectively. $Q_{\rm N}$ and $Q_{\rm B}$ are the Boltzmann distribution optimized for $P_{\rm N}$ and $P_{\rm B}$, respectively.}
\end{figure}

Figure~\ref{fig:effT}(a) shows the effective temperatures thus computed for $L=800$. It is clear that, as expected, the NASA device has a larger effective temperature, 23\% larger at $t_a=100~\mu$s, for example. This confirms the lower-noise aspect of the Burnaby device. The decrease of the effective temperature is consistent with more and more kinks being annihilated as the annealing time is increased (as seen in Fig.~\ref{fig:pl}). Indeed, the kink density for the Boltzmann distribution is
\begin{eqnarray}
\rho_{\rm kink} = \frac{1}{L}\sum_{n=0}^\infty n Q(n;\beta') = \frac{1-1/L}{1+e^{2\beta'}} ,
\label{eq:rho_kink_Boltzmann}
\end{eqnarray}
and given that we know that $\rho_{\rm kink}\propto {t_a}^{-\alpha}$, we expect $1/\beta'$ to decrease as $t_a$ increases. Moreover, a larger $\alpha$ value then corresponds to a lower value of $1/\beta'$. 
As shown in Fig.~\ref{fig:effT}(a), the effective temperature $\beta'$ obtained by equating Eq.~(\ref{eq:rho_kink_Boltzmann}) to the D-Wave data has almost the same value as the effective temperature obtained by Eq.~(\ref{eq:TN}).

Figure~\ref{fig:effT}(b) compares the empirical data and the Boltzmann distribution with the optimized effective temperature at $t_a=10~\mathrm{\mu}$s for $L=800$. Although the optimized Boltzmann distribution captures the gross shape of the kink distribution, significant differences are apparent. This is consistent with the fact that, as already mentioned in Sec.~\ref{sec:dist}, the actual distribution is close to Gaussian, as predicted by the quantum KZM theory. Thus, the deviation from the purely classical Boltzmann distribution of the kink density is to be expected. It is, however, possible that a closer agreement would be found with the quantum Boltzmann distribution obtained once quantum fluctuations freeze (at $s<1$), but this distribution too would not be Gaussian~\cite{Amin:2015qf}. We thus conclude that the kink distribution does not thermalize in accordance with equilibrium theory expectations, but is rather better described by the KZM and its generalization.

Nevertheless, Fig.~\ref{fig:effT}(b) indicates that the effective temperature obtained from our fitting  procedure can serve as a reasonable proxy for quantifying the relative overall effect of noise for a comparison between different quantum annealing devices.

\subsection{Test of a classical description by spin-vector Monte Carlo}
\label{sec:SVMC}

\begin{figure*}[t]
\includegraphics[width=\textwidth]{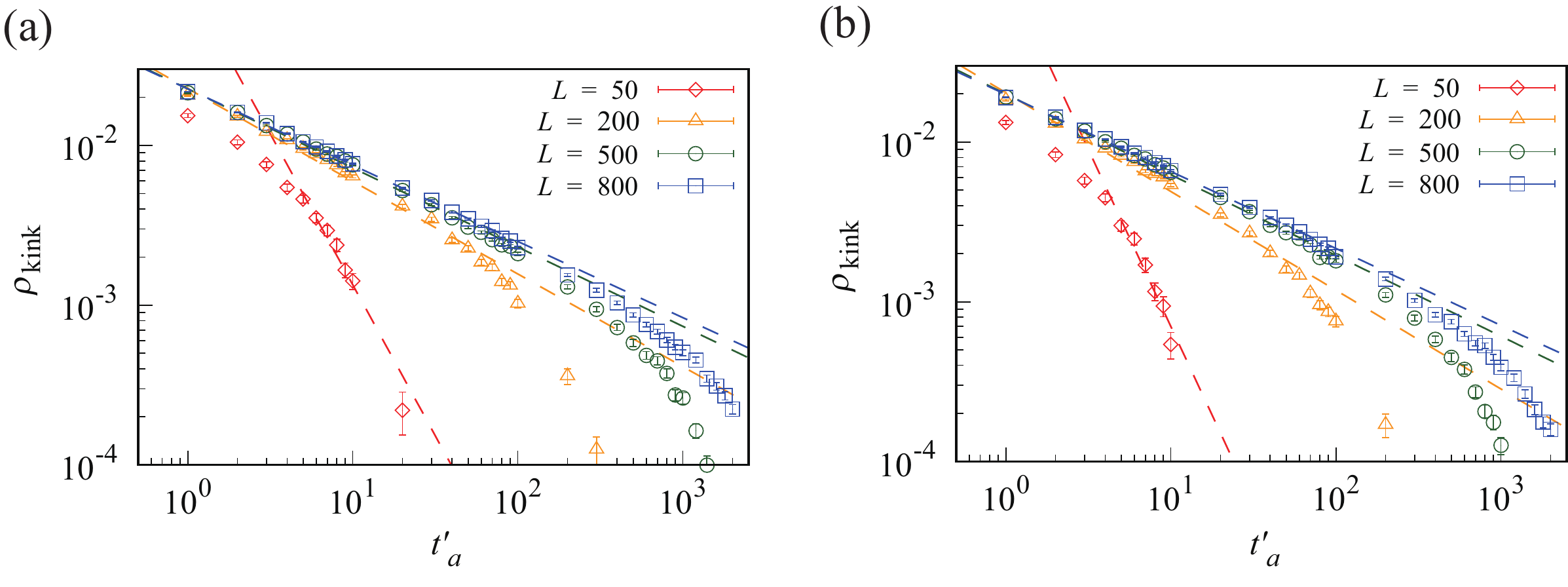}
\caption{Numerically computed kink density $\rho_{\rm kink}$ as a function of the dimensionless annealing time $t'_a$ from the SVMC model (log-log scale). The error bars indicate the 68\% confidence interval. (a) SVMC simulation results using the NASA DW2KQ annealing schedule at $T=12.1$~mK. (b) SVMC simulation results using the Burnaby DW2KQ annealing schedule at $T=13.5$~mK. Dashed straight lines are linear fits to a power-law decay. For $L=50$, the fitting range is limited from $t'_a=5$ to $t'_a=20$. Contrast with the DW2KQ results shown in Fig.~\ref{fig:pl}.}
\label{fig:svmc_pl} 
\end{figure*}

\begin{figure*}[t]
\includegraphics[width=\textwidth]{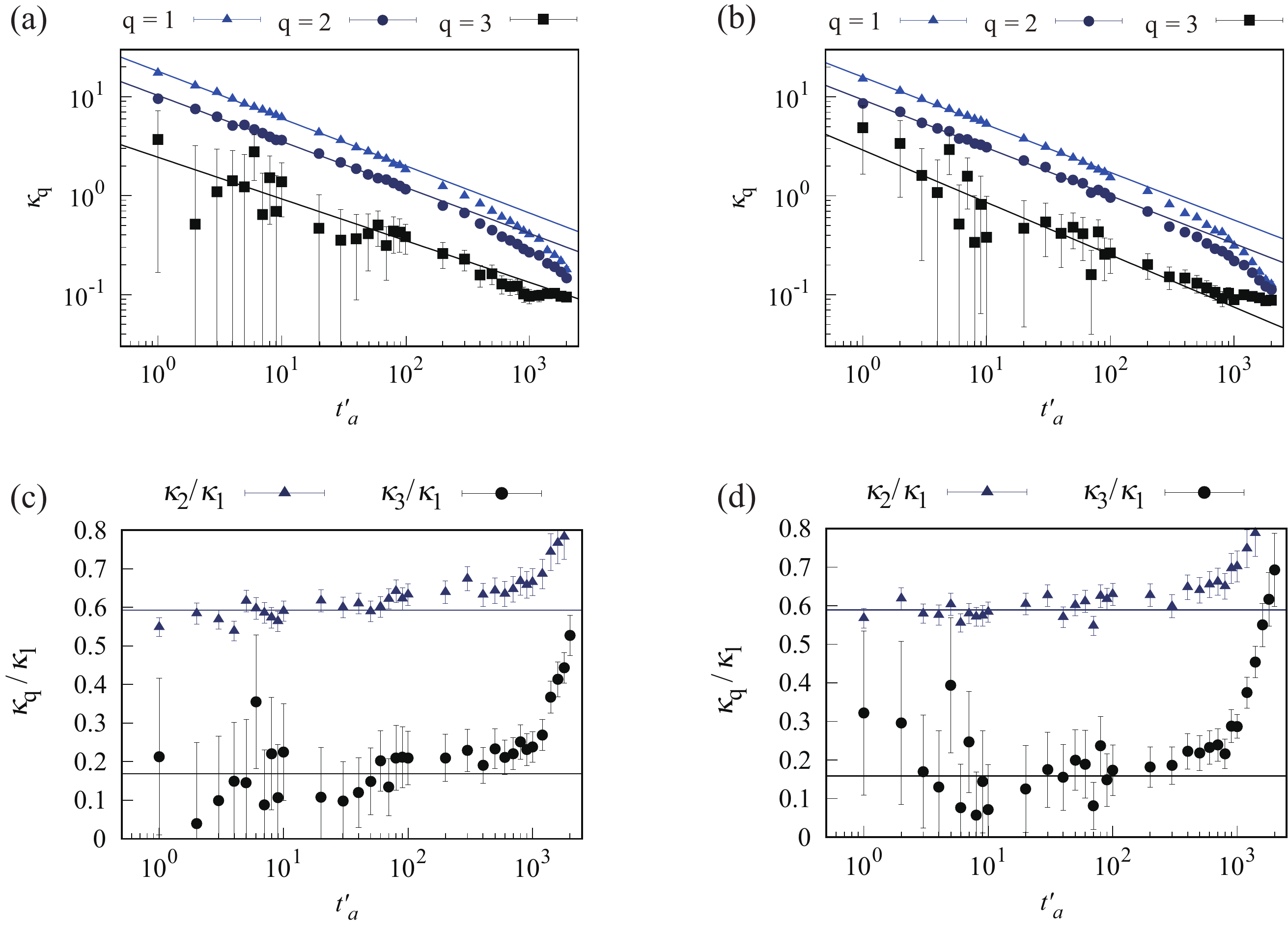}
\caption{Numerically computed cumulants of $q$th order $\kappa_{q}$ of the kink distribution. The chain length is $L=800$. Error bars indicate the $68$\% confidence interval. Solid lines are power-law fits of the cumulants from the first-order $\kappa_{1}$ to third-order $\kappa_{3}$ as functions of the dimensionless annealing time $t'_a$ with the annealing schedule of (a) the DW2KQ at NASA at $T=12.1$mK and (b) the DW2KQ in Burnaby at $T=13.5$mK. Ratios $\kappa_{2}/\kappa_1$ and $\kappa_{3}/\kappa_1$ of cumulants for the annealing schedules of (c) the DW2KQ at NASA and (d) the DW2KQ in Burnaby. Solid lines are fits to constants for the power-law decay region of the cumulants. Contrast with the DW2KQ results shown in Fig.~\ref{fig:cumulants}.}
\label{fig:svmc_cumulants} 
\end{figure*}

\begin{figure}[t]
\includegraphics[width=\columnwidth]{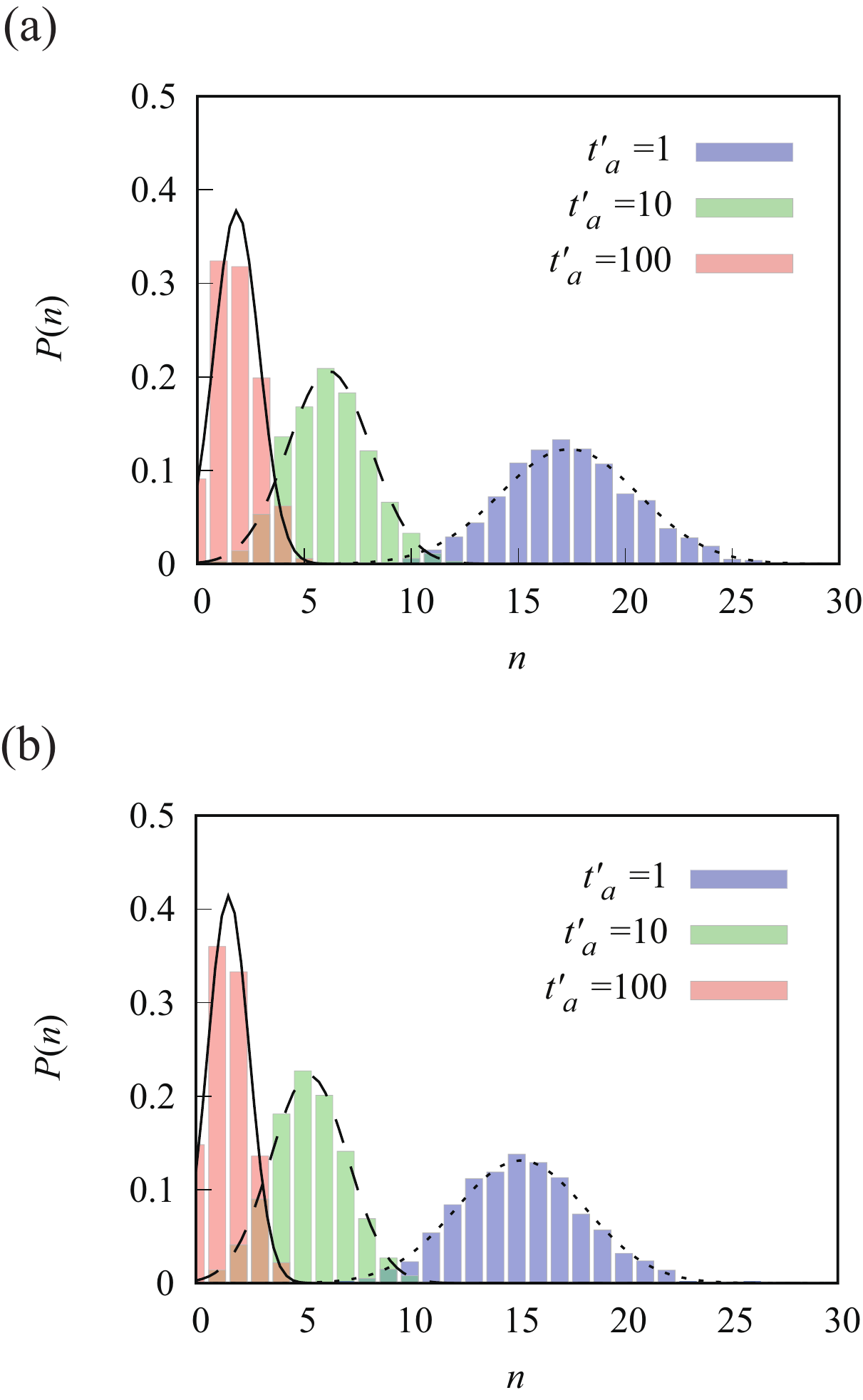}
\caption{Histograms of the number of kinks computed using the SVMC model with the annealing schedule of (a) the DW2KQ at NASA at $T=12.1$mK and (b) the DW2KQ in Burnaby at $T=13.5$mK.  The chain length is $L=800$. From right to left, the annealing times $t'_a$ are $1,10,100~\mu $s, respectively. Solid, dashed, and dotted lines are the Gaussian distributions of Eq.~\eqref{eq:nd}.Contrast with the DW2KQ results shown in Fig.~\ref{fig:dist}.}
\label{fig:dist-SVMC} 
\end{figure}

A much more stringent model than a simple Boltzmann distribution is the standard classical model of the D-Wave devices, the spin-vector Monte Carlo (SVMC) model~\cite{SSSV}. In our SVMC simulations, we replace the operators $\sigma_{i}^{x}$ and $\sigma_{i}^{z}$ by the components of a classical unit vector, $\sin{\theta_{i}}$ and $\cos{\theta_{i}}$, respectively. The Hamiltonian is therefore written as:
\footnote{The coefficient of the second term is $-A(s_n)/2$, not $A(s_n)/2$, following the convention of SVMC \cite{SSSV}.}
\begin{equation}
H=\frac{B(s_{n})}{2}\sum_{i=1}^{L-1}\cos{\theta_{i}}\cos{\theta_{i+1}}-\frac{A(s_{n})}{2}\sum_{i=1}^{L}\sin{\theta_{i}}\label{eq:svmc},
\end{equation}
where $s_{n}$ is a parameter representing time corresponding to the number of Monte Carlo steps, $n$. We choose the following parameterization of $s_n$,
\begin{equation}
s_{n}=\frac{n}{t'_aN_{0}}
\end{equation} 
where $N_0$ is the number of Monte Carlo steps necessary to reproduce the kink density observed in the D-Wave device at $1~\mathrm{\mu}$s. The dimensionless parameter $t'_a$ corresponds to the total annealing time, such that the total number of Monte Carlo steps is $n=N_0 t'_a$, and $s_n=1$ at the end of a simulation. In the present case, $N_0$ is $1000$ and $1500$ for the NASA and Burnaby devices, respectively. We use the actual NASA and Burnaby annealing schedules depicted in Fig.~\ref{fig:schedule} for comparison of the DW2KQ data with SVMC results.  We first set all local angles to $\theta_{i}=\pi/2$ and use a Metropolis move with the physical temperature of each device, $T=12.1$~mK for NASA and $T=13.5$~mK for Burnaby, and sequentially update each local state  $\theta_{i}$. After the dynamical evolution from $s=0$ to $s=1$, we project the final state to $+1$ if $0 < \theta_{i} < \pi/2 $, and  $-1$ if $\pi/2 < \theta_{i} < \pi$.  We take $1,000$ samples for each $t'_a$ for statistical analysis.

Figure~\ref{fig:svmc_pl} shows the kink density $\rho_{\rm kink}$ as a function of $t'_a$ as obtained from the SVMC simulations. The power-law scaling seen for the D-Wave data in Fig.~\ref{fig:pl} and predicted from the KZM theory is observed here as well, but only for short annealing times\footnote{The unit of time in SVMC is arbitrary and we should not directly compare the data for the same values on the horizontal axes in Figs.~\ref{fig:pl} and~\ref{fig:svmc_pl}.}.
For longer annealing times, the power law breaks down and a more rapid decay of the kink-density sets in, with the crossover point increasing with chain length $L$. This is not the case for the DW2KQ results (see Appendix~\ref{sec:Appendix2}). Moreover, the exponents extracted from the power-law regions, summarized in Table~\ref{tab:svmc_pl}, deviate substantially from the DW2KQ exponents summarized in Table~\ref{tab:pl}.  See also Fig.~\ref{fig:exponents}.
It is also shown in Appendix \ref{sec:Appendix_classical_exponent} that the critical exponent $\nu$ assumes the value $1/2$ in the SVMC model in contrast to the corresponding quantum value of $\nu=1$ for an isolated system and $\nu=0.66$ for a system coupled to a bosonic environment \cite{Werner2005}. This implies that the closeness of the decay exponent $\alpha$ of the classical SVMC model to the quantum closed-system value of $0.5$ is likely to be accidental.

\begin{table}[t]
\caption{Results from SVMC model simulations for the exponent $\alpha$ of the power-law scaling describing the decay rate of the kink density as shown in Fig.~\ref{fig:svmc_pl}. Each exponent is obtained from a fit up to the $L$-dependent crossover point seen in Fig.~\ref{fig:svmc_pl}.}
\begin{ruledtabular}
\begin{tabular}{ccc}
$L$  &  
NASA
 &
Burnaby
\\
\hline
 50  & 1.891$\pm$ 0.158 &  2.218$\pm$0.236 \\
200 & 0.580$\pm$ 0.018 & 0.618$\pm$0.021 \\
500 & 0.496$\pm$ 0.008 & 0.506$\pm$0.007 \\
800 & 0.477$\pm$ 0.005 & 0.482$\pm$0.006 \\
\end{tabular}
\end{ruledtabular}
\label{tab:svmc_pl} 
\end{table}

Another noticeable difference is that the kink density curves are all quite close, i.e. size-independent, for $L\ge 200$ for the DW2KQ results, whereas the corresponding curves tend to differ for the SVMC simulations, with the kink density decaying more slowly for larger chain lengths. 

A further test is provided by the cumulants, shown for SVMC in Fig.~\ref{fig:svmc_cumulants}. The contrast with the DW2KQ data shown in Fig.~\ref{fig:cumulants} is clear, with the constancy of the ratio seen there, as predicted from generalized KZM theory, weaker in Fig.~\ref{fig:svmc_cumulants}. We furthermore provide fits to the Gaussian distribution predicted by this theory to the SVMC simulation results in Fig.~\ref{fig:dist-SVMC}. Given the smallness of the third order cumulants, the Gaussian fits are unsurprisingly quite good, though not as good as to the DW2KQ data, shown in Fig.~\ref{fig:dist}.

Additional results for SVMC at the higher (and hence more classical) simulation temperature of $50$~mK are provided in Appendix~\ref{sec:Appendix3}. The overall trends are the same as those seen in Figs.~\ref{fig:svmc_pl}-\ref{fig:dist-SVMC}, but the agreement with the predictions of generalized KZM theory is in fact closer than for the lower temperature simulations above. In particular, agreement with the power-law decay predictions for the first cumulant extend to larger $t'_a$ values, as does the constancy of the cumulant ratios. The extracted decay exponent $\alpha$ is listed in Table \ref{tab:svmc_pl_50mK}.

\begin{table}[t]
\caption{Results from SVMC model simulations at 50~mK for the exponent $\alpha$ of the power-law scaling describing the decay rate of the kink density as shown in Fig.~\ref{fig:svmc_pl-app}. Each exponent is obtained from a fit up to the $L$-dependent crossover point seen in Fig.~\ref{fig:svmc_pl-app} of Appendix~\ref{sec:Appendix3}.}
\begin{ruledtabular}
\begin{tabular}{ccc}
$L$  &  
NASA
 &
Burnaby
\\
\hline
 50  & 1.472$\pm$ 0.134 &  1.209$\pm$0.084 \\
200 & 0.525$\pm$ 0.012 & 0.528$\pm$0.011 \\
500 & 0.450$\pm$ 0.003 & 0.455$\pm$0.004 \\
800 & 0.437$\pm$ 0.002 & 0.441$\pm$0.003 \\
\end{tabular}
\end{ruledtabular}
\label{tab:svmc_pl_50mK} 
\end{table}
The exponent $\alpha=0.44$ is closer to the experimental value 0.20 (NASA)/0.34(Burnaby) than the lower-temperature exponent $\alpha=0.48$ is, but the quantum-theoretical prediction 0.28 by the KZM is much closer to the experimental result. One noteworthy qualitative difference is that for sufficiently large annealing times, the SVMC kink density deviates downward from the power-law (fewer kinks), while in the D-Wave case it deviates upward (more kinks). 

\begin{figure}[h]
\includegraphics[width=\columnwidth]{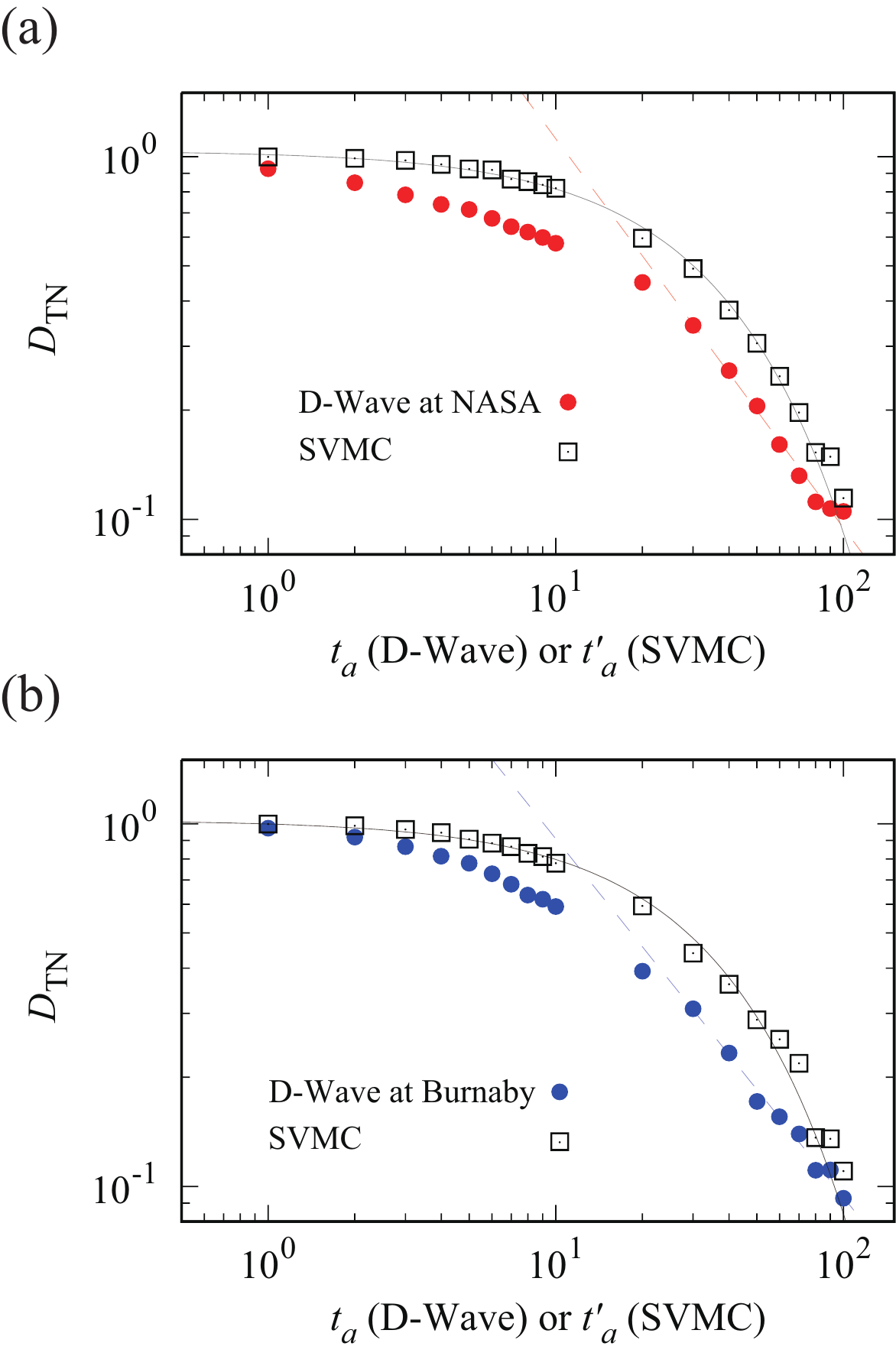}
\caption{Trace-norm distance between the kink density of the D-Wave device or SVMC simulations and the Boltzmann distribution with the same $\beta'$, chosen such that the distances are minimized at $t_a=100~\mathrm{\mu}$s. The chain length is $L=800$. The solid black lines are exponentials, $D_{\rm}\propto e^{-\gamma t_a}$, fitted to the SVMC data. The dotted lines are polynomials, $D_{\rm}\propto t_a^{-\tau}$, fitted to the D-Wave data for $t_a\in[30,100]~\mathrm{\mu}$s using (a) the NASA and (b) Burnaby annealing schedules.}
\label{fig:svmc_TN} 
\end{figure}

Finally, we also analyzed the D-Wave and SVMC data by computing their trace-norm distance from the Boltzmann distribution. Here the goal is to test the prediction of the adiabatic theorem for open quantum systems, that this distance decreases following a power law as a function of time $t_a$ for sufficiently large $t_a$~\cite{Venuti2016}. Figures~\ref{fig:svmc_TN} (a) and (b) show the result for $L=800$. We  computed the trace-norm distance according to Eq.~\eqref{eq:TN}, which uses the optimized Boltzmann distribution $Q\left(n;\beta'\right)$ in Eq.~\eqref{eq:Gibbs}, and the kink distributions of D-Wave and the SVMC simulations as empirical distributions $P(n)$. We fixed $\beta'$ of $Q\left(n;\beta'\right)$ to that already obtained at $t_a=100~\mu$s (D-Wave) or $t'_a=100$ (SVMC) because we are interested in how the trace-norm distance approaches the Boltzmann distribution at this annealing time, the largest reliable value available to us (see Appendix~\ref{sec:Appendix1}). It is seen from Fig.~\ref{fig:svmc_TN} that the decrease of the trace-norm distance of SVMC fits an exponential (solid line), while the behavior of the D-Wave is closer to a power law for sufficiently large $t_a$ although the difference is not large.
Thus it is reasonable to conclude that the D-Wave results are in closer agreement with the adiabatic theorem for open quantum systems than the classical SVMC simulation results. 

Given all the discrepancies we have found, it is reasonable to conclude that the SVMC model does not explain the behavior of the DW2KQ devices reported here.

\section{Discussion}
\label{sec:discussion}

We have reported on extensive  experiments for the one-dimensional transverse-field Ising model performed using the NASA and low-noise Burnaby  DW2KQ devices. We demonstrated that the kink density decays in a power law $t_a^{-\alpha}$ with the annealing time $t_a$, in qualitative agreement with the theoretical prediction of the Kibble-Zurek mechanism (KZM). In more detail, we found that the exponent $\alpha$ describing the rate of power-law decay differs from the KZM prediction derived under the assumption of an isolated, closed quantum system. The difference between the theoretical value of $\alpha=0.5$ (which, coincidentally, is close to the outcome of the classical SVMC model as well), and the empirical values of $\alpha=0.20$ (NASA) and $\alpha=0.34$ (Burnaby), can be understood to a first approximation by modeling the coupling of the system to a bosonic environment with an Ohmic spectral density, which reduces the theoretical value to $\alpha=0.28$. Although it is difficult to quantitatively explain the remaining discrepancy, it is reasonable to suppose that it originates in other factors,
which lead to a non-universal value of the exponent, such as parameter control errors, and transient effects due to short annealing times. Indeed, the larger exponent (a faster decrease of the kink density) of the Burnaby device than the NASA device is consistent with the lower-noise characteristics of the former.

The lower noise aspect of the Burnaby device was further verified in our study by computing the effective temperature of the best-fit Boltzmann distribution at the end of the anneal, which shows that the effective temperature is about 23\% higher on the NASA device. Note that this effective temperature reflects the combined effects of the dilution refrigerator temperature (which is in fact slightly higher for the Burnaby device) and a wide range of other noise sources including coupling to the environment and control errors.

Related work was reported by Gardas \textit{et al}.~\cite{Gardas2018} on the previous generation D-Wave 2X devices, at Los Alamos National Laboratory and in Burnaby. They also found a power-law decay of the kink density but the value of the exponent $\alpha$ depended strongly on experimental conditions such as the choice of the device and the sign of interactions (ferromagnetic or antiferromagnetic). The values reported range from $\alpha=0.24$ to $\alpha=1.31$, and they left the explanation for further work after listing several possible options. In contrast, our work quite definitively establishes that quantum simulation using the newer DW2KQ devices is capable of demonstrating and probing the KZM and its generalization, in particular using the lower-noise version in Burnaby with the~\texttt{DW\_2000Q\_5} solver.

Other closely related work is the recent experiment probing the two-dimensional transverse-field Ising model on the square lattice by Weinberg \textit{et al}.~\cite{Weinberg2019}, which demonstrated non-monotonicity in the kink density as a function of the annealing time $t_a$.  In the short annealing time regime, the kink density decreases as a function of $t_a$, as in our case.
The value of the exponent they found in this shorter time range, $\alpha=0.74$, is close to the theoretical value for an isolated system in two dimensions, $\alpha=0.77$.  In contrast, for long annealing times, the kink density increases as $t_a$ increases. This kind of behavior is often referred to as anti-Kibble-Zurek scaling and can result from environmentally-induced heating~\cite{Dutta16,puebla2019akzm}.  Weinberg \textit{et al}. also attribute this latter behavior to the effects of noise, including thermal fluctuations.  Indeed, numerical calculations for the one-dimensional system shown in Fig.~\ref{fig:iTEBD} and presented in Ref.~\cite{Weinberg2019} as well as in Refs.~\cite{Patane08,Suzuki2019} show that the kink density can be non-monotonic if the temperature is finite.  A possible reason that our experiment did not find such non-monotonicity in the range $t_a\le 100~\mathrm{\mu}$s is that this annealing time is too short for the temperature effects to appear in the one-dimensional problem. Our data in the range of very long annealing times up to $2000~\mathrm{\mu}$s show non-monotonicity in some cases and possibly reflect thermal and other deviations from the ideal quantum simulation, as discussed in Appendix~\ref{sec:Appendix1}.  We excluded this time range from our analysis since the data appears unstable with large uncertainties. In the short time region of the two-dimensional experiment of Ref.~\cite{Weinberg2019} where the KZM is likely to apply, the system seems to be much less susceptible to noise and the exponent $\alpha$ is close to the theoretical value of an ideal, isolated system as mentioned above. These observations suggest that how noise affects the system behavior strongly depends on the problem type as well as on the annealing time range.

We further investigated the distribution of the kink density at the end of the anneal, which encodes signatures of universality beyond the original predictions of KZM theory. We found agreement with the theoretical prediction that the ratio of the second and higher order cumulants $\kappa_q~(q\ge 2)$ to the first order cumulant $\kappa_1$ is independent of the annealing time $t_a$~\cite{DelCampo2018}. We also found very good agreement with the prediction of a Gaussian distribution of the kink density. This agreement with a quantum theory constructed for a closed, isolated system suggests that these are robust features that remain largely intact even in the presence of coupling to an environment. 
Extensive numerical computations using iTEBD with QUAPI support this conclusion.  In the classical KZM, it is indeed the case that the constancy of cumulant ratios is a robust feature \cite{GomezRuiz19b}, but a generalization of this classical theory to the quantum case is non-trivial.

Given the history of challenges via classical models to experiments involving the D-Wave devices~\cite{Smolin,SSSV,SSSV-comment}, and their rebuttals~\cite{comment-SS,Albash:2014if,q-sig,q-sig2,DWave-entanglement,Albash:2015pd,Boixo2016,Job:2017aa}, we tested whether classical models alternatively explain the experimental data as well. We first tried a simple fit to a Boltzmann distribution but did not find satisfactory agreement. We also tried the standard classical limit of the D-Wave device, the spin-vector Monte Carlo (SVMC) model~\cite{SSSV} and found that the quantum theory of the generalized KZM provides better qualitative as well as quantitative agreement.

The two DW2KQ devices we tested therefore serve to a good approximation as quantum simulators of the one-dimensional transverse-field Ising model under the influence of a dephasing Ohmic bosonic environment. These bosons do not represent thermal effects because we have observed neither an approach of the kink density to a constant, nor a non-monotonic behavior of the kink density as a function of annealing time (over the annealing time range where we have confidence in the reliability of the data).
Instead, the bosons possibly correspond to dynamical fluctuations of the normal current flowing through Josephson junctions~\cite{Caldeira1983}.  It is a difficult but interesting future direction of work to identify the nature of these fluctuations and to try to find a way to reduce them for better agreement with the closed quantum system limit.

\section{Conclusion}
\label{sec:summary}
In the wake of a phase transition, topological defects form. The Kibble-Zurek mechanism predicts that the density of defects scales as a universal power-law with the  time scale used to cross the transition. We have shown that this prediction can be tested on quantum annealers by using them for analog quantum simulation, i.e., as a test-bed for non-equilibrium statistical mechanics. Specifically, our work has tested the Kibble-Zurek mechanism in the one-dimensional transverse-field Ising model. Our analysis of the quantum annealer data shows that the behavior of the devices is consistent with the implementation of this model coupled to a bosonic environment.  Our work thus provides experimental evidence of universal Kibble-Zurek scaling in an open quantum system.

By probing the full counting statistics of topological defects (kinks), we furthermore established signatures of universality beyond the original prediction of the Kibble-Zurek scaling, which is focused on the average kink number. In particular, we found that the power-law scaling with the annealing time of the average kink number is shared by its variance and the third cumulant. Our experimental and numerical results thus indicate that the universal scaling recently predicted for the cumulants of the kink number distribution in an isolated quantum critical system also holds under open-system quantum dynamics,

\begin{acknowledgments}
The research of YB, DL, and HN is based upon work supported by the Office of the Director of National Intelligence (ODNI), Intelligence Advanced Research Projects Activity (IARPA), via the U.S. Army Research Office contract W911NF-17-C-0050. The views and conclusions contained herein are those of the authors and should not be interpreted as necessarily representing the official policies or endorsements, either expressed or implied, of the ODNI, IARPA, or the U.S. Government. The U.S. Government is authorized to reproduce and distribute reprints for Governmental purposes notwithstanding any copyright annotation thereon.
\end{acknowledgments}

\appendix

\begin{figure*}[t]
\includegraphics[width=\textwidth]{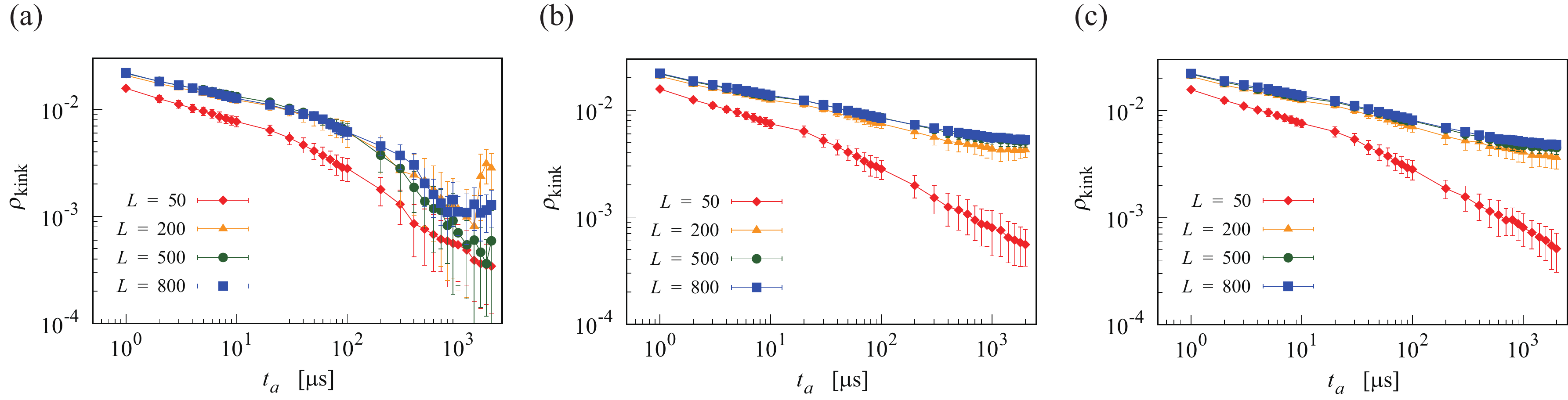}
\caption{\label{fig:append_gauge}Kink density as a function of the annealing time with different interactions on the DW2KQ at NASA. The error bars indicate the 68\% confidence interval. (a) Ferromagnetic interactions ($J=-1)$; (b) Antiferromagnetic interactions ($J=1)$; (c) Random gauge, where we randomly select $L/2$ sites $i$ and flip the sign of interactions between qubits $i-1,i$ and  $i,i+1$.}
\end{figure*}

\begin{figure*}[t]
\includegraphics[width=\textwidth]{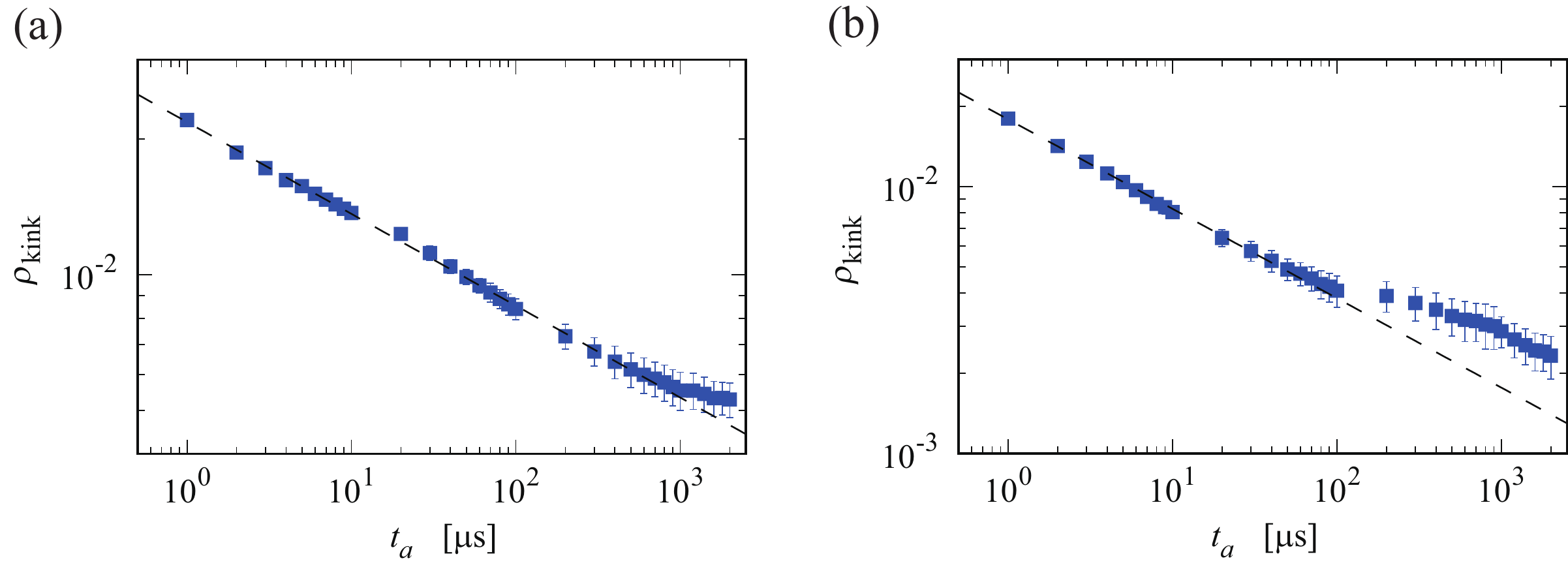}
\caption{Kink density for $L=800$ as a function of the annealing time from $t_a=1~\mathrm{\mu}$s to $t_a=2000~\mathrm{\mu}$s on (a) the NASA DW2KQ device and (b) the Burnaby DW2KQ device. In both cases we used antiferromagnetic interactions. The error bars indicate the 68\% confidence interval. The dashed lines are power law fits to the data up to 100~$\mathrm{\mu}$s.}
\label{fig:append_long}
\end{figure*}

\begin{figure*}[t]
\includegraphics[width=\textwidth]{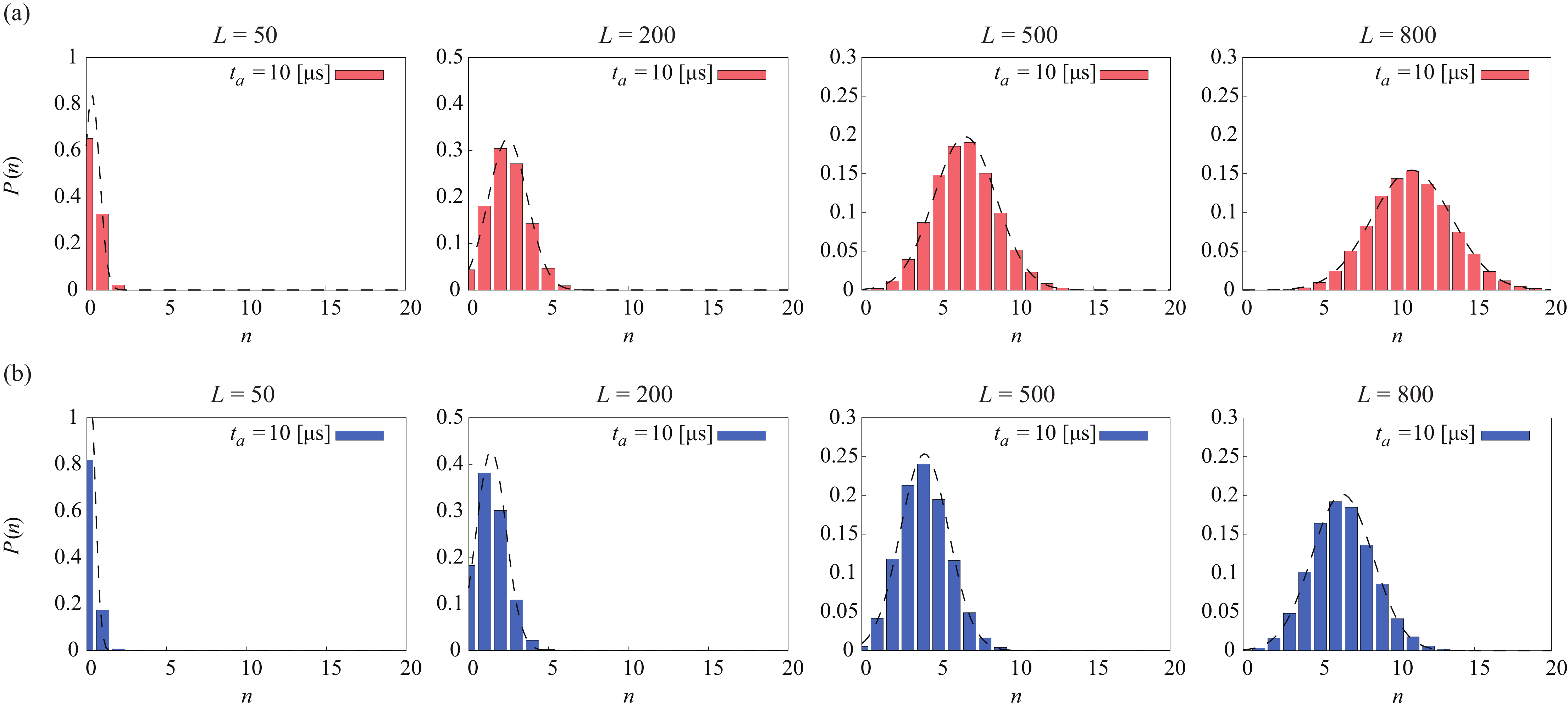}
\caption{Kink distribution at $t_a=10~\mu s$ for different chain lengths $L=50, 200, 500, 800$ (left to right) from the DW2KQ devices at (a) NASA and (b) Burnaby. The dashed lines are the theoretical prediction of Eq.~\eqref{eq:nd}.}
\label{fig:append_dist}
\end{figure*}

\section{Dependence of the kink density on the interaction type and annealing time}
\label{sec:Appendix1}

Figure~\ref{fig:append_gauge} shows the kink density obtained for three types of interactions up to $t_a=2000~\mathrm{\mu}$s on the NASA DW2KQ device. Figure~\ref{fig:append_gauge}(a) displays the case of ferromagnetic interactions ($J=-1$), (b) antiferromagnetic ($J=1$), and (c) random gauge. In the latter case we first set all the interactions to be ferromagnetic, randomly choose $L/2$ qubits, and change the sign of their interactions with their two nearest neighbors. In one dimension with free boundaries, these three cases are theoretically equivalent and the kink density should behave identically.

Figures~\ref{fig:append_gauge}(a), (b), and (c) clearly indicate that the differences are small for the short time regime $t_a\le 100~\mathrm{\mu}$s. Beyond this regime, marked deviations emerge in the ferromagnetic case, whereas the antiferromagnetic and random gauge data remain close even up to the longest annealing time $t_a=2000~\mathrm{\mu}$s. This difference may imply the presence of a systematic bias toward ferromagnetic states in the D-Wave devices, which becomes prominent at larger annealing times.
Antiferromagnetic and random-gauge interactions may be interpreted to have caused cancellations of such a bias.  For these reasons we choose to use the data from antiferromagnetic interactions in the main text.

Additional data displayed in Fig.~\ref{fig:append_long} show that the kink density obeys a power-law decay very accurately up to $t_a=100~\mathrm{\mu}$s on both the NASA and the Burnaby DW2KQ devices with antiferromagnetic interactions and chain length $L=800$.  Beyond $t_a=100~\mathrm{\mu}$s, deviations from a power law become apparent. This data set, along with Fig.~\ref{fig:append_gauge}, motivated us to use the empirical data only for $t_a\le 100~\mathrm{\mu}$s, in order to eliminate artifacts other than the coupling to bosonic environment.

\section{Kink distribution for different chain lengths}
\label{sec:Appendix2}

Figure~\ref{fig:append_dist} supplements Fig.~\ref{fig:dist} by showing histograms of the kink density for different chain lengths on the two different devices, (a) NASA and (b) Burnaby, at $t_a=10\mu$s.  Dashed lines are the Gaussian distribution of Eq.~\eqref{eq:nd}. The data are well described by this distribution in all cases.

\section{Critical exponent $\nu$ for the SVMC model}
\label{sec:Appendix_classical_exponent}
We show in this Appendix that the critical exponent $\nu$ is 1/2 for the one-dimensional ferromagnetic SVMC model.
The Hamiltonian is
\begin{align}
    H=-J\sum_{j=1}^N \sin\theta_j \sin\theta_{j+1}-\Gamma \sum_{j=1}^N \cos \theta_j.
\end{align}
We have exchanged $\sin\theta_j$ and $\cos \theta_j$ from the conventional notation of Eq.~(\ref{eq:svmc}) for later convenience. A periodic boundary is assumed.

Since we are interested in how the system behaves as we decrease $\Gamma$ from a very large value (where the system is in the paramagnetic phase with $\theta_j\approx 0~\forall j$) toward a transition point,  it is reasonable to expand the Hamiltonian to quadratic order as
\begin{align}
    H=-J\sum_j \theta_j \theta_{j+1}+\frac{\Gamma}{2}\sum_j \theta_j^2,
\end{align}
where we have ignored a constant term. Let us check if the paramagnetic state is stable by Fourier transformation,
\begin{align}
    \theta_j=\frac{1}{N}\sum_{k=1}^N e^{-i2\pi kj/N}\phi_k
\end{align}
as
\begin{align}
    H=\frac{1}{2}\sum_k A_k |\phi_k|^2,~ A_k=\frac{1}{N}\Big(\Gamma -2J\cos \frac{2\pi k}{N}\Big).
    \label{HA}
\end{align}
It is observed that $\phi_k=0~\forall k$ is the stable state (ground state) configuration for $\Gamma >2J$, and a second-order phase transition exists at $\Gamma_c=2J$.

The correlation is given by
\begin{align}
    G(r)&=\langle \sin\theta_0 \sin\theta_r\rangle \approx \frac{1}{N}\sum_{l=1}^N \langle \theta_l \theta_{l+r}\rangle \nonumber\\
   & =\frac{1}{N^2}\sum_k e^{i2\pi k r/N}\langle |\phi_k|^2 \rangle, \label{GA}
\end{align}
where the angular brackets $\langle \cdots \rangle$ stand for the statistical-mechanical average with respect to the Hamiltonian (\ref{HA}). We have averaged over $l$ using translation symmetry. 
Since the Hamiltonian (\ref{HA}) represents independent Gaussian fields, we easily find
\begin{align}
    \langle |\phi_k|^2 \rangle
=\frac{1}{\beta A_k}=\frac{k_B T N}{\Gamma -2J\cos (2\pi k/N)}.
\end{align}
Thus the correlation function (\ref{GA}) becomes
\begin{align}
    G(r)=\frac{k_B T}{N}\sum_k e^{i2\pi k r/N} \frac{1}{\Gamma -2J\cos (2\pi k/N)}.
\end{align}
The behavior as $r\gg 1$ is evaluated as
\begin{align}
    G(r)
    = \frac{k_B T}{2\pi} \int_{0}^{\infty} e^{iyr}\frac{1}{\Gamma -2J +Jy^2}
    dy \propto  e^{-r/\xi},
\end{align}
where
\begin{align}
    \xi =\sqrt{\frac{J}{\Gamma -2J}}.
\end{align}
Thus the exponent is $\nu =1/2$. Notice that the temperature is kept small but finite. If $T=0$ exactly, no fluctuations exist classically and the spin configuration is fixed to $\theta_j=0$ in the paramagnetic phase. 

\section{SVMC results at $T=50$~mK}
\label{sec:Appendix3}

\begin{figure*}[t]
\includegraphics[width=.9\textwidth]{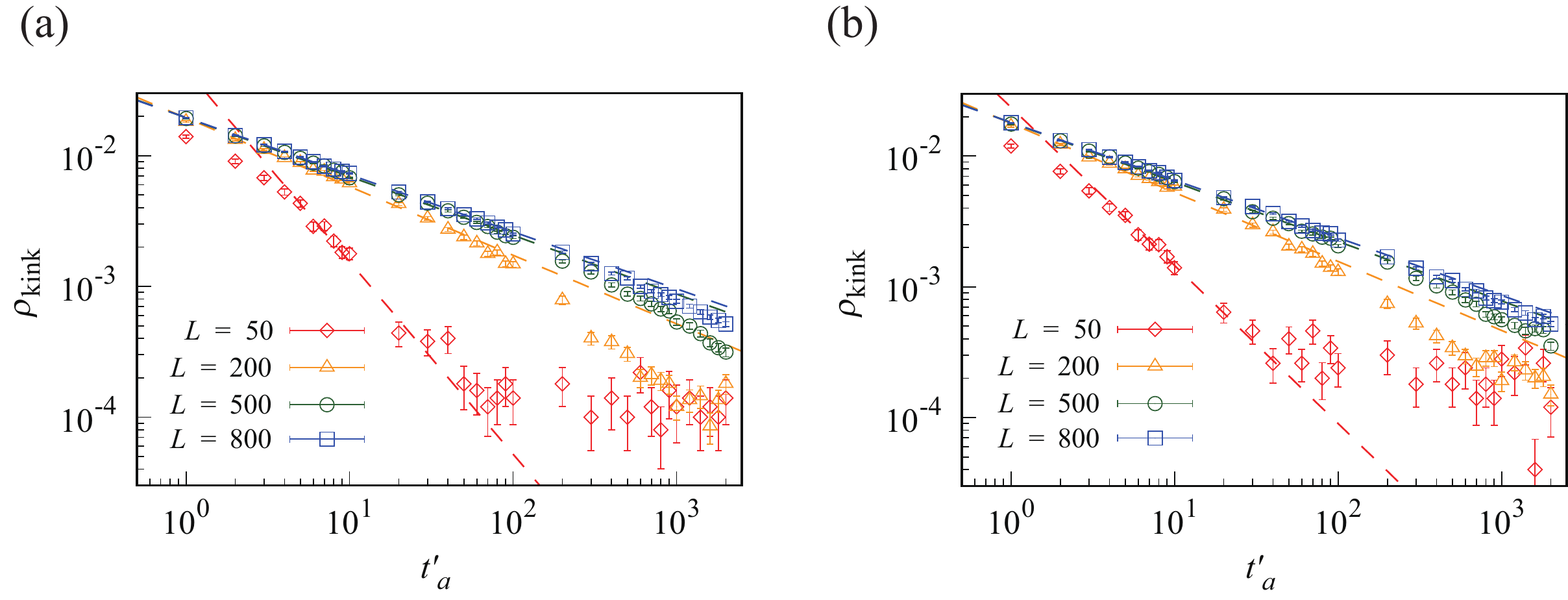}
\caption{Same as Fig.~\ref{fig:svmc_pl} but for $T=50$~mK: numerically computed kink density $\rho_{\rm kink}$ as a function of the dimensionless annealing time $t'_a$ from the SVMC model (log-log scale). The error bars indicate the 68\% confidence interval. (a) SVMC simulation results using the NASA DW2KQ annealing schedule. (b) SVMC simulation results using the Burnaby DW2KQ annealing schedule. Dashed straight lines are linear fits to a power-law decay. For $L=50$, the fitting range is limited from $t'_a=5$ to $t'_a=20$.}
\label{fig:svmc_pl-app} 
\end{figure*}

\begin{figure*}[t]
\includegraphics[width=.9\textwidth]{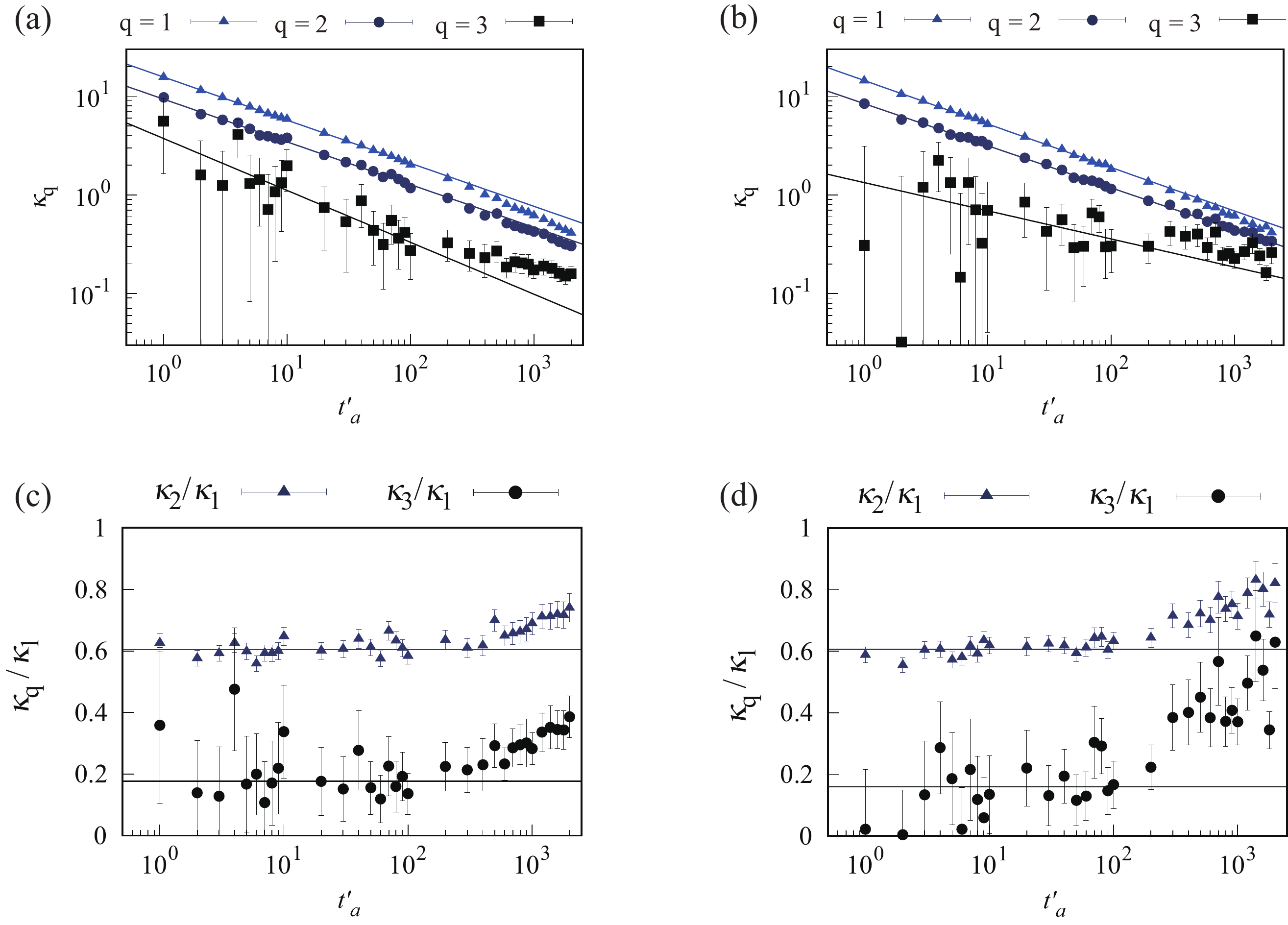}
\caption{Same as Fig.~\ref{fig:svmc_cumulants} but for $T=50$~mK: numerically computed cumulants of $q$th order $\kappa_{q}$ of the kink distribution. The chain length is $L=800$. Error bars indicate the $68$\% confidence interval. Solid lines are power-law fits of the cumulants from the first-order $\kappa_{1}$ to third-order $\kappa_{3}$ as functions of the dimensionless annealing time $t'_a$ with the annealing schedule of (a) the DW2KQ at NASA and (b) the DW2KQ in Burnaby. Ratios $\kappa_{2}/\kappa_1$ and $\kappa_{3}/\kappa_1$ of cumulants for the annealing schedules of (c) the DW2KQ at NASA and (d) the DW2KQ in Burnaby. Solid lines are fits to constants for the power-law decay region of the cumulants.}
\label{fig:svmc_cumulants-app} 
\end{figure*}

\begin{figure}[t]
\includegraphics[width=.9\columnwidth]{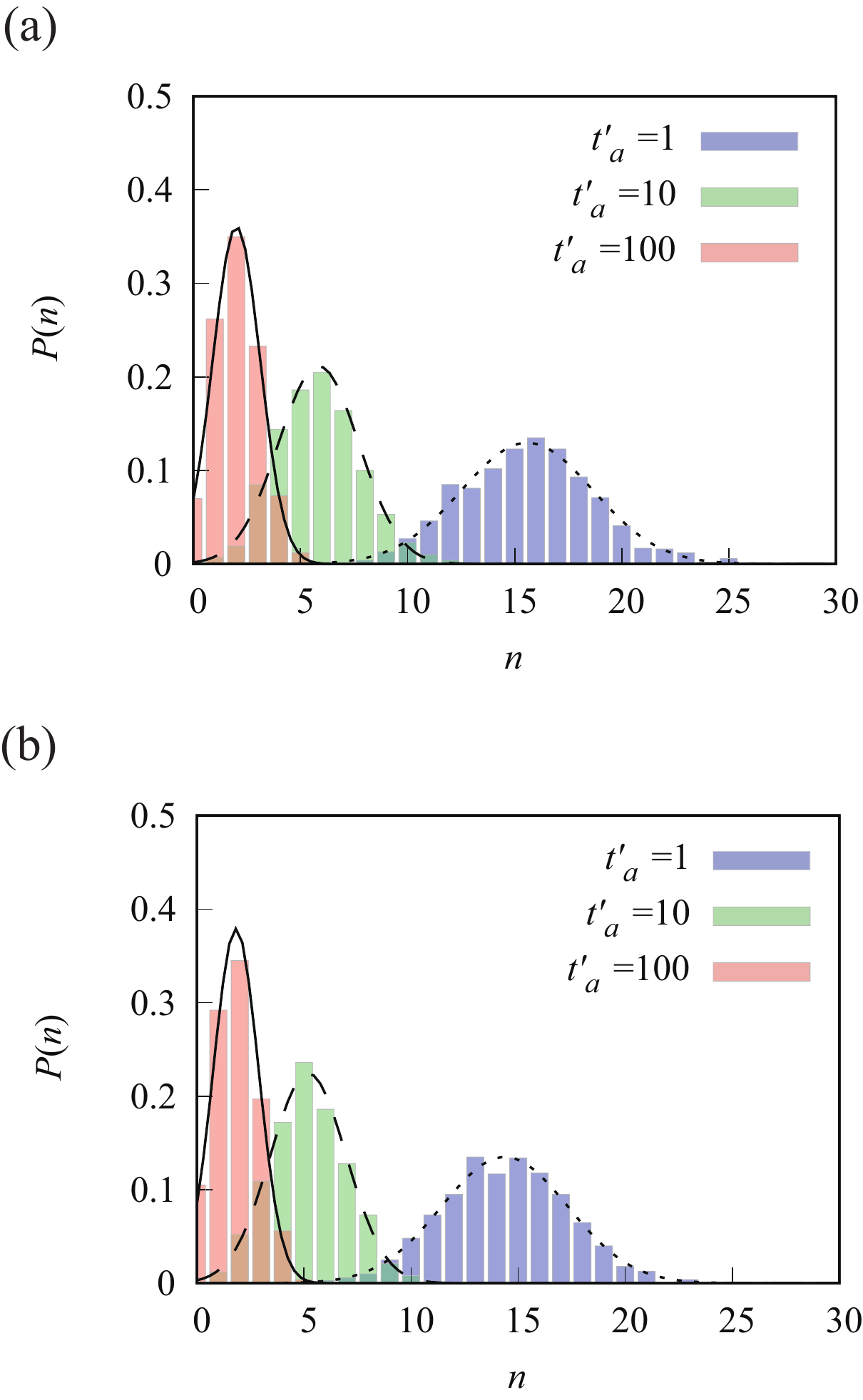}
\caption{Same as Fig.~\ref{fig:dist-SVMC} but for $T=50$~mK: histograms of the number of kinks computed using the SVMC model with the annealing schedule of (a) the DW2KQ at NASA and (b) the DW2KQ in Burnaby.  The chain length is $L=800$. From right to left, the annealing times $t'_a$ are $1,10,100~\mu $s, respectively. Solid, dashed, and dotted lines are the Gaussian distributions of Eq.~\eqref{eq:nd}.}
\label{fig:dist-SVMC-app} 
\end{figure}

In the main text we provided our SVMC results at the dilution refrigerator temperatures of the NASA and Burnaby DW2KQ devices. Here we provide additional SVMC results computed at a higher simulation temperature of $50$~mK in Figs.~\ref{fig:svmc_pl-app}, \ref{fig:svmc_cumulants-app}, and \ref{fig:dist-SVMC-app}. The $\alpha$ values corresponding to Fig.~\ref{fig:svmc_pl-app} are reported in Table~\ref{tab:svmc_pl_50mK}. These values are somewhat closer to the DW2KQ values than those for the dilution refrigerator temperatures, and we also note that the qualitative behavior of the kink density curves seen in Fig.~\ref{fig:svmc_pl-app} is more like the DW2KQ results seen in Fig.~\ref{fig:pl}, in that the curves for the two largest $L$ values now nearly overlap. In these respects the warmer SVMC model is a closer match to the DW2KQ data than at the dilution refrigerator temperatures.

\clearpage

\bibliography{mainKZM.bib}

\begin{thebibliography}{95}
\bibitem [{kin()}]{kink2defect}%
  \BibitemOpen
  \href@noop {} {}\bibinfo {note} {We use both terms
  interchangeably.}\BibitemShut {Stop}%
\end{thebibliography}

\end{document}